\documentclass[aps,prb,preprint,amsmath,amssymb,floatfix]{revtex4-2}
\usepackage{graphicx}
\usepackage{amsfonts}
\usepackage{mathrsfs}
\usepackage{multirow}
\usepackage[version=4]{mhchem}
\usepackage{siunitx}
\usepackage{subcaption}
\usepackage{textcomp}
\usepackage{gensymb}
\usepackage{hyperref}
\usepackage{booktabs}
\usepackage{enumitem}

\begin{document}

\title{Spin splitting without symmetry: a nearly compensated ferrimagnet and the origin of altermagnetism}

\author{Jaekyung Jang}
\affiliation{Center for Advanced Computation, Korea Institute for Advanced Study, Seoul 02455, Republic of Korea}

\author{Joo Yull Rhee}
\email{rheejy@skku.edu}
\thanks{Corresponding author}
\affiliation{Department of Physics, Sungkyunkwan University, Suwon 16419, Republic of Korea}

\begin{abstract}
The spin splitting of collinear altermagnets is usually attributed to the crystal symmetry relating the opposite-spin sublattices --- a rotation or mirror in place of the translation or inversion of a conventional antiferromagnet.  We show that this symmetry organizes the splitting but does not produce it; its origin is the anisotropic arrangement of the magnetic orbitals and their ligands, which also fixes the symmetry itself.  Two results establish this. A triclinic ($P1$) \ce{Mn} oxide, a fully compensated ferrimagnet whose two inequivalent sublattices are related by no symmetry, is spin split throughout the Brillouin zone; repositioning the same ligands, atom for atom, to restore an inversion centre relating the two metal sites collapses the splitting to a residual two orders of magnitude smaller, so the splitting follows from the arrangement rather than the symmetry.  Symmetry, we show formally, can only ever reverse or forbid a splitting that already exists: it cannot create one between sublattices an arrangement has left electronically identical, nor where the underlying orbitals carry no anisotropy to arrange in the first place.  Organizing the analysis through an operation-resolved symmetry hierarchy and reading the crystal classes as a single axis, the enforced nodal planes fall from two to one to zero across the orthorhombic, monoclinic, and triclinic classes while the arrangement-generated splitting persists to the base.  The altermagnet and the fully compensated ferrimagnet are thus the two ends of one axis --- the same splitting, organized by symmetry at one end and unorganized at the other --- and the search for compensated spin-split magnets is a search over orbital arrangements rather than symmetry labels.
\end{abstract}

\maketitle

\section{Introduction}
\label{sec:introduction}

A spin splitting of the electronic bands --- an energy difference, defined in Eq. \eqref{eq:Delta_definition}, between the two spin channels --- is the defining electronic feature of a magnetically ordered metal or insulator. In a ferromagnet the splitting is uniform in sign and accompanies a finite net moment. In a conventional antiferromagnet it is absent: the net moment vanishes and the two spin channels remain degenerate at every momentum, protected by a symmetry that maps one spin sublattice onto the other. The recognition that a third possibility exists --- a collinear magnet with vanishing net moment whose bands are nonetheless spin split in a momentum-dependent way --- has driven the rapid development of altermagnetism \cite{Smejkal2022a,Naka2021,Okugawa2018}. Here the opposite-spin sublattices are related not by a translation or inversion, which would force degeneracy, but by a rotation or mirror, which permits a splitting that alternates in sign across the Brillouin zone and vanishes on symmetry-fixed nodal planes, giving the characteristic $d$-, $g$-, or $i$-wave textures.

The prevailing account of this phenomenon is framed almost entirely in terms of symmetry: one identifies the operation relating the spin sublattices, and from it reads off whether spin splitting is allowed and what angular form it takes. This account is correct as far as it goes, but it carries an implicit assumption alongside its explicit content. Explicitly, it states which symmetry organizes the splitting into which nodal structure. Implicitly, it is read as though that symmetry \emph{produces} the splitting --- as though, absent the special rotation or mirror, there would be no splitting to organize. The two readings are rarely separated, and separating them is the purpose of this work. The question is whether symmetry is the origin of the spin splitting or only its organizer, and the two are answered differently by the same body of evidence.

The distinction becomes sharp at a case the symmetry account sets aside. A collinear magnet with vanishing net moment need not be an antiferromagnet. If the two spin sublattices are crystallographically inequivalent --- carrying moments that are unequal in magnitude and merely happen to cancel --- the state is a fully compensated ferrimagnet, connected by no symmetry at all \cite{Guo2025,Dong2025}. Such states have been of recent interest, largely in engineered low-dimensional settings where the compensation is imposed by construction. Their existence raises a question that the altermagnetic framework, built on the presence of a relating symmetry, does not address: what happens to the spin splitting when \emph{no} symmetry relates the sublattices? A vanishing net moment has been taken to imply either an antiferromagnet, degenerate, or, given the right symmetry, an altermagnet, split; the fully compensated ferrimagnet, with no relating symmetry, falls outside both, and its band structure has gone largely unexamined.

The question of whether spin splitting requires a sublattice-relating crystal symmetry has been approached from several directions, each loosening that requirement while stopping short of removing it. Interaction-driven mechanisms show that a splitting can appear in a lattice without crystallographic sublattice anisotropy, when electronic correlations drive an orbital ordering that lowers the effective symmetry to an altermagnetic one \cite{Leeb2024}; and altermagnetism has been shown to form in amorphous solids that lack global rotational symmetry, through a common point-group symmetry generated locally by spin and orbital ordering \cite{Ornellas2026}. In both, the splitting is still organized by a symmetry relating the opposite-spin sublattices --- one revived by the orbital ordering in the first case, assembled from local symmetries in the second. In both, moreover, the orbital ordering or arrangement must appear before that symmetry can do anything, so that their own constructions can be read as placing the arrangement first. That reading, however, is ours: neither work states it, and each continues to present the symmetry as what defines the resulting state. The step we take is to make the implication explicit and to follow it to its end --- that the arrangement, not the symmetry, is what generates the splitting, and that when the arrangement revives no relating symmetry the result is not an altermagnet but a fully compensated ferrimagnet. Whether this is said explicitly is not a matter of emphasis: it is the difference between treating the splitting as a consequence of symmetry and understanding where it actually originates. That is the case we isolate.

The closest account of where the splitting comes from is that of \v{S}mejkal \textit{et al.}\ \cite{Smejkal2022b}. There the size and momentum dependence of the splitting follow the \emph{electric crystal potential}, defined as the internal potential of the nonmagnetic phase, as given for example by a density-functional calculation without magnetic order. The term is theirs, and we use it only in that sense. A potential of the nonmagnetic phase contains the positions of all ions, and so it contains the shape of the ligand cage around each magnetic atom. We read the anisotropy it carries as the anisotropic shape of the magnetic orbital that this cage produces. They show this for \ce{KRu4O8}, where the orbitals of the two sublattices have anisotropy axes rotated relative to each other, and describe the mechanism for materials with orbital degeneracy in the nonmagnetic phase. That description joins two conditions under one name. We state them separately.

The first condition is that the orbital has a direction at all. An ideal oxygen octahedron treats its three axes alike. It is not spherical, but it is isotropic: it does not distinguish one axis from another. A direction appears only when the cage distorts, when the distances between the magnetic ion and its ligands become unequal and the bonds are no longer related by the cubic operations. For the half-filled Mn $3d$ shell of \ce{MnO}, this direction resides mainly in the Mn--O hybridization. The condition belongs to a single atom. It needs no symmetry relating different atoms, and it is the source of the splitting: without it, nothing can differ between the spin channels.

The second condition is that the distorted orbitals of the magnetic atoms are not placed alike. If every magnetic atom holds its orbital in the same orientation, a translation relates them and the splitting vanishes. When the orientations differ, by rotation or tilt, the up and down channels come to differ. This second condition triggers the splitting. Only here does the altermagnetic symmetry enter, because the operations that relate the sublattices are fixed by how the orientations are placed.

The roles are not interchangeable. With the second condition met and the first missing, nothing happens, because an isotropic orbital has no orientation to place differently. With the first met and the second missing, nothing happens either, and symmetry can even cancel a splitting that the first would allow. In \ce{KCuF3} every Cu sits in a Jahn--Teller-distorted octahedron, yet A-type order gives zero splitting, because a translation relates the two sublattices (Sec.~\ref{sec:admissibility}).

We test this in a triclinic ($P1$) MnO whose only spatial operation is the identity. The bands are spin split throughout the Brillouin zone, with no symmetry to relate the sublattices or to enforce any nodal plane. Repositioning the same ligands, atom for atom, to restore an inversion centre between the two metal sites collapses the splitting to a residual two orders of magnitude smaller. The two calculations differ only in the ligand positions, not in the magnetic decoration or in the ligand species and number. The splitting is present where no symmetry can account for it, and it collapses when only the arrangement changes. We use \emph{arrangement} for the two conditions together: the anisotropy of the single orbital and the different placement of that anisotropy on the two sublattices. We prefer it to the established \emph{orbital ordering}. Ordering names a state, present or absent, and does not say why an ordered state should split the bands. Arrangement names the disposition from which the splitting follows. The same arrangement also determines the symmetry itself, by building the operation that relates the sublattices where the bare metal sites have none, or, on its removal, by reviving one that had been suppressed. Symmetry is thus not the origin of the splitting but a feature the arrangement sets, alongside the splitting it generates.

This is not only what the calculation shows but what the standard picture, examined closely, already requires: a symmetry relating opposite-spin sublattices can only interchange bands that are already split, and cannot separate degenerate ones, so it organizes a splitting it did not create.  The triclinic case and this observation together fix the causal order.

This reorders the usual picture. We develop it through an operation-resolved hierarchy that separates the admissibility of splitting, the network of sublattice-relating operations, and the enforced nodal geometry, and we apply it to orthorhombic \ce{LaMnO3}, to octahedrally tilted vetoes in \ce{KV2Se2O} and \ce{KCuF3}, and to minimal monoclinic and triclinic Mn oxides. Read across the crystal classes as a single axis, the number of symmetry-enforced nodal planes falls from two to one to zero, while the arrangement-generated splitting persists to the triclinic base. The altermagnet and the fully compensated ferrimagnet emerge as the two ends of this axis --- the same arrangement-generated splitting, organized by symmetry into a definite wave character at one end and unorganized at the other. Symmetry classifies the outcome; the arrangement is its origin. The remainder of the paper is organized as follows. Section~\ref{sec:symmetry_hierarchy} sets out the hierarchy; Sec.~\ref{sec:permissive_generative} distinguishes the permissive role of symmetry from the generative role of the arrangement; Sec.~\ref{sec:admissibility} treats the tilting-controlled veto; Secs.~\ref{sec:bands}--\ref{sec:enforced_accidental} develop the orthorhombic and monoclinic cases; Sec.~\ref{sec:triclinic} presents the triclinic ferrimagnet; and Secs.~\ref{sec:discussion}--\ref{sec:conclusion} draw the conclusions together. Further computational details, additional crystal structures, band structures, and magnetic-moment data are given in the Supplemental Material (SM) \cite{suppl}.

\section{The symmetry hierarchy of collinear altermagnetic nodal structure}
\label{sec:symmetry_hierarchy}
This section presents the operation-resolved hierarchy used in the remainder of this work.  Throughout, we consider collinear magnetic order in the nonrelativistic, spin--orbit-free limit, so that the two spin channels can be treated independently.  The spin splitting of band $n$ is
\begin{equation}
    \Delta_n(\mathbf{k})
    =
    E_{n\uparrow}(\mathbf{k})
    -
    E_{n\downarrow}(\mathbf{k}).
    \label{eq:Delta_definition}
\end{equation}
For an altermagnet, in spite of vanishing net moment, $\Delta_n(\mathbf{k})$ is generically nonzero, but the two spin sectors remain related by spatial symmetry at different momenta.

Identification rules for altermagnetism have been given previously \cite{Smejkal2022b}, in the form of conditions to be checked: an even number of magnetic atoms, the absence of an inversion centre relating opposite-spin sites, a connecting rotation, and the spin group fixed by a classification algorithm.  These are consistent with the present construction, but are not given as an ordered sequence --- the veto condition appears alongside the others rather than as a gate that must be cleared before they become meaningful.  Section~\ref{sec:admissibility} shows why the order matters, and the hierarchy developed here fixes it explicitly, resolved in each case by direct structural observation rather than by the full spin-group classification algorithm.

The same limitation recurs whenever splitting is judged by a symmetry or motif condition alone: such a condition can rule a texture in or out, but --- as the Yuan--Zunger motif-pair reading of C-type \ce{LaMnO3} (Sec.~\ref{subsec:setting_convention}) and the loop-current condition for $p$-wave magnetism (Sec.~\ref{sec:discussion}) will make explicit --- it never addresses what sets the magnitude of the splitting it permits.

The symmetry ingredients entering this description are established in spin-group and spin-Laue formulations of collinear magnetism \cite{Smejkal2022a,Smejkal2022b,Naka2021}, in the pre-altermagnetic analysis of opposite-spin site operations in perovskites \cite{Okugawa2018}, and in related design principles \cite{Wei2024}.  Our purpose is not to add a symmetry criterion but to separate the distinct physical roles these ingredients play and to place them in an explicit ordered sequence.  Three questions must be answered in turn.  First, is nonrelativistic spin splitting admissible at all?  Second, if it is, which spatial operations relate the opposite-spin sectors across momentum space?  Third, which invariant manifolds of those operations enforce zeros of the splitting and thereby fix the nodal geometry? These questions define the three levels developed below.  The section closes with two consequences of the hierarchy that are used repeatedly in the material analysis: a plane-supplier principle, Eq. \eqref{eq:plane_supplier}, relating the number of enforced nodal planes to the spin-exchange operations that possess invariant planes, and a resulting organization of collinear altermagnets by crystal class.

\subsection{Spin-preserving and spin-exchanging operations}
\label{subsec:PX_general}
Let $O$ be a spatial operation compatible with a collinear magnetic configuration.  Its action on the magnetic sublattices is classified by whether it maps a magnetic site onto a site of the same or of the opposite spin.  We introduce the spin-sublattice parity
\begin{equation*}
    \eta(O)
    =
    \begin{cases}
        +1, & O \text{ preserves the spin sublattice},\\
        -1, & O \text{ exchanges the spin sublattices},
    \end{cases}
    \label{eq:eta_general}
\end{equation*}
and, for compactness,
\begin{equation*}
    p\equiv +1,
    \qquad
    e\equiv -1.
    \label{eq:PX_notation}
\end{equation*}
The symbols $p$ and $e$ refer to the magnetic action of a spatial operation, not to additional crystallographic elements.  A single spatial operation of the parent crystal may act as $p$ in one magnetic decoration and as $e$ in another; this reassignment is central to the material analysis of Secs.~\ref{sec:admissibility}--\ref{sec:enforced_accidental}.

For symmetry-related bands the momentum-space relation is
\begin{equation}
    \Delta_n(O\mathbf{k})
    =
    \eta(O)\,\Delta_n(\mathbf{k}),
    \label{eq:PX_general}
\end{equation}
so that a spin-preserving operation gives $\Delta_n(O\mathbf{k})=+\Delta_n(\mathbf{k})$ and a spin-exchanging operation gives $\Delta_n(O\mathbf{k})=-\Delta_n(\mathbf{k})$, or equivalently
\begin{equation}
    E_{n\uparrow}(O\mathbf{k})
    =
    E_{n\downarrow}(\mathbf{k})
    \label{eq:X_energy_general}
\end{equation}
for appropriately matched bands.  The spatial operation does not itself reverse spin; its action on the magnetic decoration determines whether it connects same-spin or opposite-spin sites.

The parity is multiplicative,
\begin{equation}
    \eta(O_2O_1)
    =
    \eta(O_2)\,\eta(O_1),
    \label{eq:eta_multiplicative_general}
\end{equation}
so that
\begin{equation*}
    p\times p= p,
    \qquad
    p\times e= e,
    \qquad
    e\times e= p.
    \label{eq:PX_multiplication}
\end{equation*}
This is the operation-resolved form of the familiar subgroup--coset structure of collinear spin groups \cite{Smejkal2022a,Smejkal2022b}.  If $\mathcal G$ is the relevant spatial symmetry group and $\mathcal H$ its same-spin subgroup, the opposite-spin operations form the complementary coset,
\begin{equation}
    \mathcal G
    =
    \mathcal H
    \cup
    A\mathcal H ,
    \label{eq:coset_general}
\end{equation}
with $A$ any representative spin-exchanging operation.  What matters for the present analysis is how this structure is used.  A single $C_2$, $C_4$, or $C_6$ operation can label the angular character of a state compactly, but it does not display the full spatial content of $A\mathcal H$.  Different members of the coset generally act differently on momentum space, possess different invariant manifolds, and therefore carry independent geometrical information.  Retaining the complete set $A\mathcal H$ rather than a single representative is what allows the nodal geometry to be reconstructed, as shown at Level~III and applied throughout the material sections.

\subsection{Level I: admissibility --- the veto}
\label{subsec:level1}

The first level concerns the existence of the splitting itself. Before any angular structure is considered, magnetic symmetry must permit $\Delta_n(\mathbf{k})$ to be finite at generic $\mathbf{k}$.

Nonrelativistic spin splitting is forbidden when a symmetry operation connects the opposite-spin sublattices while leaving the crystal momentum unchanged.  A pure translation $T_{\mathbf t}$ that exchanges the two spin sublattices,
\begin{equation}
    T_{\mathbf t}:\ \uparrow\leftrightarrow\downarrow,
    \label{eq:X_translation}
\end{equation}
changes a Bloch state only by a phase, so Eq.~\eqref{eq:X_energy_general} becomes a same-$\mathbf{k}$ statement,
\begin{equation}
    E_{n\uparrow}(\mathbf{k})
    =
    E_{n\downarrow}(\mathbf{k}),
    \qquad\Longrightarrow\qquad
    \Delta_n(\mathbf{k})=0
    \label{eq:translation_degeneracy}
\end{equation}
throughout the Brillouin zone.  An inversion centre relating opposite-spin sublattices acts in the same way: combined with the even parity of the nonrelativistic band energy, $E_{n\sigma}(-\mathbf{k})=E_{n\sigma}(\mathbf{k})$, it likewise forces same-$\mathbf{k}$ degeneracy.  Either equivalence restores the two spin sectors at every momentum and eliminates the splitting globally.

This is qualitatively different from a spin-exchanging rotation, reflection, glide, or screw whose point action carries a generic $\mathbf{k}$ to a distinct momentum $O\mathbf{k}$.  Such an operation imposes $E_{n\uparrow}(O\mathbf{k})=E_{n\downarrow}(\mathbf{k})$ but does not require degeneracy at the same generic $\mathbf{k}$, so finite splitting remains allowed.

The admissibility level is therefore a \emph{veto}: it asks whether a global same-$\mathbf{k}$ degeneracy-restoring equivalence is present,
\begin{equation}
\boxed{
\text{Does a translation or inversion connect }\uparrow\leftrightarrow\downarrow
\text{ at fixed }\mathbf{k}?
}
\label{eq:admissibility_question}
\end{equation}
If it does, the answer terminates the analysis: $\Delta_n(\mathbf{k})\equiv0$, and any additional spin-exchanging operations the structure may contain merely reduce to $0=-0$ and organize no finite texture.  If no such equivalence exists, finite splitting is admissible and the analysis proceeds.  This is the general form of the degeneracy-restoring exclusion emphasized in material-design and site-symmetry approaches \cite{Wei2024,Alam2026}; pure translations and inversion-related cases are its two elementary realizations.  The logical priority matters: the operations that organize a momentum-dependent texture become relevant only after this veto has been cleared, a point made concrete by the magnetic-order and polytype selection in Sec.~\ref{sec:admissibility}.

\subsection{Level II: positive existence and the spin-exchange network}
\label{subsec:level2}

Once the veto is cleared, altermagnetism additionally requires a \emph{positive} condition: a spin-exchanging spatial operation must actually connect the sublattices,
\begin{equation}
    \exists\,g\in\mathcal G:\quad \eta(g)=-1, \quad g\mathbf{k}\neq\mathbf{k}\ \text{(generic }\mathbf{k}).
    \label{eq:positive_existence}
\end{equation}
Absent such an operation the two sectors are unrelated by symmetry and the state is a ferromagnet or ferrimagnet rather than a compensated altermagnet.  A useful corollary follows immediately: a nontrivial spin-exchanging point operation with a moving action on $\mathbf{k}$ requires at least a twofold rotation or a mirror, so a collinear altermagnet must belong at least to a monoclinic point group.  Triclinic symmetry, whose only nontrivial operation is inversion, cannot supply Eq.~\eqref{eq:positive_existence} and yields either a ferrimagnet or, if inversion is spin-exchanging, a veto-suppressed conventional antiferromagnet.

Beyond mere existence, the object that controls the texture is the complete opposite-spin sector $A\mathcal H$ of Eq.~\eqref{eq:coset_general}, not a single representative.  A representative relation
\begin{equation*}
    \Delta_n(C_N\mathbf{k}) = -\Delta_n(\mathbf{k})
    \label{eq:CN_exchange}
\end{equation*}
fixes the angular character --- momenta connected by $C_N$ carry opposite signs --- and is often sufficient for classification.  But different elements of $A\mathcal H$ relate different momentum pairs and possess different invariant lines or planes.  The representative establishes the transformation $\mathbf{k}\to O\mathbf{k}$, $\Delta_n\to-\Delta_n$; the full set additionally determines \emph{where} the sign changes are forced through zero.  The second level is therefore a network problem,
\begin{equation}
\boxed{
\text{Which elements of }A\mathcal H\text{ relate the opposite-spin sectors?}
}
\label{eq:network_question}
\end{equation}
It does not decide whether altermagnetism exists --- that is settled at Level~I together with Eq.~\eqref{eq:positive_existence} --- but fixes the complete spatial organization of the allowed splitting.  In high-symmetry crystals the distinction is easily overlooked, because repeated application of a single $C_4$ or $C_6$ generates several angular sectors on its own.  It becomes transparent in lower-symmetry altermagnets, where a $d$-wave texture need not arise from a crystallographic $C_4$ axis at all but from distinct orthogonal spin-exchanging operations, as in orthorhombic \ce{LaMnO3} below.

\subsection{Level III: nodal geometry and the plane-supplier principle}
\label{subsec:level3}

The third level converts the network into geometrical constraints on $\Delta_n(\mathbf{k})$.  Let $g_i$ be a spin-exchanging operation and $\mathcal N_i$ a momentum manifold invariant under its point action modulo a reciprocal-lattice vector,
\begin{equation*}
    g_i\mathbf{k} = \mathbf{k}+\mathbf G,\qquad\mathbf{k}\in\mathcal N_i .
    \label{eq:g_invariant}
\end{equation*}
On $\mathcal N_i$ the relation $\Delta_n(g_i\mathbf{k})=-\Delta_n(\mathbf{k})$ reduces to $\Delta_n=-\Delta_n$, requiring
\begin{equation}
    \Delta_n(\mathbf{k})=0,\qquad \mathbf{k}\in\mathcal N_i .
    \label{eq:X_enforced_node}
\end{equation}
The invariant manifold of a spin-exchanging operation is therefore a symmetry-enforced nodal manifold.  No such conclusion holds for a spin-preserving operation, whose invariant manifold yields the identity $\Delta_n=+\Delta_n$:
\begin{equation*}
\boxed{
\begin{aligned}
\text{an } e \text{ operation on its invariant manifold}
&\ \Longrightarrow\
\Delta_n=0,\\
\text{a } p \text{ operation on its invariant manifold}
&\ \not\Longrightarrow\
\Delta_n=0.
\end{aligned}}
\label{eq:PX_nodal_rule}
\end{equation*}
The collection of $e$-invariant manifolds constitutes the enforced nodal surfaces; the regions between them are connected by the spin-exchanging operations and carry alternating signs of $\Delta_n$.

This is shown by the composition $C_2\cdot {\bar 1}$.  A spin-exchanging $C_2$ alone gives only a nodal line; composing it with inversion converts the invariant axis into an invariant plane, so that the line becomes an enforced nodal plane.  A plane, in other words, requires a mirror, and the mirror can be supplied in two ways.  In a centrosymmetric crystal it is a crystallographic element, so $C_2\cdot {\bar 1}$ is a mirror and the plane is enforced by the space group directly.  In a chiral or low-symmetry crystal whose only relevant symmetry is the bare rotation, with no crystallographic inversion available, the even parity of the nonrelativistic, spin--orbit-free band energy, $E_{n\sigma}(-\mathbf{k})=E_{n\sigma}(\mathbf{k})$, plays the role of ${\bar 1}$ at the level of the spin Hamiltonian, so the same mirror is defined at the Hamiltonian level and the plane is enforced just as it would be by a crystallographic one.  Either way, it is a mirror --- crystallographic or Hamiltonian-supplied --- that a nodal plane requires; a rotation or line-supplying operation alone is not sufficient.  The two mechanisms are unified by the plane-supplier count: what matters for an enforced plane is whether a spin-exchanging operation with a two-dimensional invariant set is available, whether it originates in the crystallographic group or in the parity of the nonrelativistic Hamiltonian.

A further point concerns where such a plane appears, once supplied.  A mirror of this kind acts as a pure reflection of a single momentum component, $k_\beta\to-k_\beta$, leaving the other two components untouched.  Because crystal momentum is defined only modulo a reciprocal-lattice vector, the invariance condition $k_\beta\equiv-k_\beta$ has, on the periodic Brillouin zone, not one but exactly two solutions: $k_\beta=0$, the zone centre, and $k_\beta=1/2$, the zone boundary.  Both loci are equally invariant under the mirror, and Eq.~\eqref{eq:X_enforced_node} applies to either with the same force: wherever the mirror is spin exchanging, the zone-boundary section is an enforced nodal plane exactly as the zone-centre one is, a parallel copy of the same plane supplier rather than a second, independent one.  This does not add to the count in Eq.~\eqref{eq:plane_supplier}, which counts operations, not their parallel copies; it does mean that an enforced plane can appear as a group rather than singly.  The grouping is specific to the one momentum direction the mirror acts on: for either of the other two directions, which the operation leaves untouched, no such plane is implied at all, at the centre or at the boundary, and any nodal behaviour there is accidental.  Whether a given zone-boundary section turns out to be enforced is therefore not a separate question but the same plane-supplier question asked at $k_\beta=1/2$ instead of $k_\beta=0$; the material sections below make direct use of this in both the orthorhombic and monoclinic cases.

For the four-sector $d$-wave case relevant below, two intersecting enforced planes divide an appropriate momentum section into sectors with alternating signs $+,-,+,-$, and their orientation fixes the orientation of the $d$-wave texture.  A change of magnetic decoration can reorient the texture by changing which parent operations act as $e$, even with the crystallographic structure held fixed.

Finally, not every zero of $\Delta_n(\mathbf{k})$ is enforced. Additional nodal points, lines, or surfaces can arise from the detailed electronic structure --- hopping amplitudes, orbital hybridization, particular dispersions --- and are constrained only to occur in a symmetry-compatible pattern, without symmetry fixing their location or even their existence:
\begin{equation}
\begin{aligned}
\text{enforced:}&
\text{spin-exchange symmetry}\Rightarrow
\text{existence and location of the node},\\
\text{accidental:}&
\text{electronic structure}\Rightarrow
\text{node},\ 
\text{symmetry constrains only its shape}.
\end{aligned}
\label{eq:plane_supplier}
\end{equation}
This distinction is not merely formal: Sec.~\ref{sec:enforced_accidental} demonstrates, using band-resolved nodal maps of a minimal monoclinic model, that accidental nodal surfaces can differ in geometry from band to band and can even be present for one band and absent for another in the same plane --- accidental in both location and existence.

\subsection{Crystal-class tiers}
\label{subsec:tiers}

The plane-supplier principle, Eq. \eqref{eq:plane_supplier}, organizes collinear altermagnets by the number of enforced nodal planes their point group can supply, and hence by crystal class.

\paragraph{Triclinic ($\bar1$, $1$).}
No spin-exchanging operation with a moving action on $\mathbf{k}$ is available.  Equation~\eqref{eq:positive_existence} fails, and the compensated state is a ferrimagnet, or --- if inversion is spin-exchanging --- a veto-suppressed conventional antiferromagnet. Altermagnetism is excluded.

\paragraph{Monoclinic ($2$, $m$).}
A single spin-exchanging $C_2$ or mirror clears the veto and satisfies Eq.~\eqref{eq:positive_existence}.  The point group supplies exactly one operation with a two-dimensional invariant set, so Eq.~\eqref{eq:plane_supplier} gives \emph{one} enforced nodal plane (the mirror plane for $m$, or the plane obtained by promoting the $C_2$ axis through the even parity of the nonrelativistic energy, as above).  A four-sector $d$-wave texture then requires a \emph{second} nodal surface that is not symmetry-enforced.  Monoclinic altermagnets are therefore the minimal setting in which enforced and accidental nodal surfaces coexist; the $P2$ and $Pm$ models of Secs.~\ref{sec:minimal_model} and \ref{sec:enforced_accidental} realize this case, in which the accidental surface can be identified separately from the enforced plane.

The centrosymmetric class $2/m$ adds nothing to this tier.  Its inversion is either spin exchanging, in which case it connects opposite-spin sublattices and activates the Level-I veto (Sec.~\ref{subsec:level1}), giving a conventional antiferromagnet rather than an altermagnet; or it is spin preserving, in which case it joins the same-spin subgroup and, through $m=\bar1\cdot C_2$ and the multiplicativity of $\eta$, forces $C_2$ and $m$ to share the same parity, so no new spin-exchange operation and no new invariant plane appears beyond the single one already counted.  The altermagnetic content of the monoclinic tier is therefore exhausted by the $2$ and $m$ classes: a $2/m$ configuration either reduces to a conventional antiferromagnet with no spin splitting or reproduces the band structure of the corresponding $2$ or $m$ case, so it adds no independent altermagnetic information and requires no separate calculation.  The $P2$ and $Pm$ \ce{MnO} models of Secs.~\ref{sec:minimal_model} and \ref{sec:enforced_accidental} thus realize the full altermagnetic content of the tier.

\paragraph{Orthorhombic and above ($D_{2h}$ and higher).}
The orthorhombic point group $D_{2h}$ contains three mutually orthogonal twofold axes $C_{2a},C_{2b},C_{2c}$ and, in a centrosymmetric setting, the three corresponding mirrors.  The mirrors are the plane suppliers of Eq.~\eqref{eq:plane_supplier}, since it is their point parts that leave a momentum plane invariant; we write the cyclic algebra on the $C_{2a},C_{2b},C_{2c}$ labels only as a compact idealization.  In a nonsymmorphic space group these operations are realized as glides and screws carrying fractional translations, which leave the enforced planes unchanged (the spin splitting is gauge invariant) but enter as Bloch phases at the zone boundary; the concrete $Pbnm$ realization is given for \ce{LaMnO3} in Sec.~\ref{sec:bands}.  Their spin parities are constrained by the multiplicativity Eq.~\eqref{eq:eta_multiplicative_general}, which for the closed relation $C_{2a}C_{2b}=C_{2c}$ gives the cyclic algebra
\begin{equation}
    \eta(C_{2a})\,\eta(C_{2b}) = \eta(C_{2c}),\qquad\eta(C_{2a})\,\eta(C_{2b})\,\eta(C_{2c})=+1 .
    \label{eq:cyclic_triple}
\end{equation}
The triple product is fixed to $+1$, so the admissible parity assignments are exactly the permutations of $(p,e,e)$ together with $(p,p,p)$. The all-$p$ case is the non-altermagnetic configuration in which no axis exchanges spin; each of the three $(p,e,e)$ permutations selects one spin-preserving axis and \emph{two} spin-exchanging ones, hence \emph{two} enforced nodal planes and a genuine four-sector \textit{d}-wave texture, with no accidental surface \emph{required} for it. Accidental surfaces are not thereby excluded: the electronic structure may add them regardless of how many enforced planes the symmetry supplies, here or in still higher-symmetry ($C_4$, $C_6$) altermagnets. What the high-symmetry case removes is only the \emph{need} for an accidental surface, not its possibility. Orthorhombic \ce{LaMnO3}, whose A-, G-, and C-type orders realize the three $(p,e,e)$ permutations within one fixed parent crystal, is the natural vehicle for this tier and is analyzed in Sec.~\ref{sec:bands}.

The three tiers are summarized in Table~\ref{tab:tiers}.  Together with the ordered levels of Eqs.~\eqref{eq:admissibility_question}, \eqref{eq:network_question}, and~\eqref{eq:plane_supplier}, they provide the complete diagnostic sequence
\begin{equation*}
\boxed{
\text{veto}
\xrightarrow{\text{cleared}}
\text{positive existence}
\xrightarrow{\ \text{yes}\ }
\left\{
\begin{array}{l}
\text{enforced planes}\ (\#=\text{plane-suppliers})\\[2pt]
\text{accidental surfaces}
\end{array}
\right.
}
\label{eq:full_sequence}
\end{equation*}
used to analyze the material realizations in the following sections.

\begin{table}[tb]
\caption{Crystal-class tiers of collinear nonrelativistic
altermagnetism, classified by the number of symmetry-enforced nodal
planes ($n$) supplied by the point group through
Eq.~\eqref{eq:plane_supplier}.}
\label{tab:tiers}
\begin{tabular}{@{}llcl@{}}
\toprule
Crystal class & Spin-exchange operation & $n$ & Outcome \\
\midrule
Triclinic & none (only $I$) & 0 & ferri- / conventional AFM \\
Monoclinic & one $C_2$ or $m$ & 1 & AM; 2nd surface accidental \\
Orthorhombic $+$ & two orthogonal $e$ & $\ge 2$ & AM; enforced $d$-wave \\
\bottomrule
\end{tabular}
\end{table}

\section{Permissive symmetry versus generative arrangement}
\label{sec:permissive_generative}

The hierarchy of Sec.~\ref{sec:symmetry_hierarchy} determines where the spin splitting must vanish and how it transforms between symmetry-related momenta. It does not, and cannot, fix the magnitude of the splitting. This distinction is often blurred by the shorthand that ``symmetry generates the splitting,'' and clarifying it is a prerequisite for interpreting both the material comparisons below and the broader altermagnetic literature. We therefore separate two logically distinct factors: the \emph{arrangement} --- the anisotropic orbitals and their non-isotropic placement between sublattices, defined in Sec.~\ref{sec:introduction}, which together generate the magnitude of any splitting --- and the spatial symmetry that fixes only the nodal geometry.

\subsection{What symmetry does and does not supply}
\label{subsec:permissive_role}

Consider a spin-exchanging operation $g$ with $\Delta_n(g\mathbf{k})=-\Delta_n(\mathbf{k})$. Equation~\eqref{eq:X_enforced_node} guarantees $\Delta_n=0$ on the invariant manifold of $g$, and Eq.~\eqref{eq:PX_general} relates $\Delta_n(\mathbf{k})$ to $\Delta_n(g\mathbf{k})$ elsewhere. Both statements are homogeneous in $\Delta_n$: multiplying the splitting by any constant leaves them satisfied. Symmetry therefore constrains the \emph{form}
\begin{equation*}
    \Delta_n(\mathbf{k})
    =
    \lambda_n(\mathbf{k})\,f_{\rm sym}(\mathbf{k}),
    \label{eq:form_amplitude}
\end{equation*}
where $f_{\rm sym}$ is the symmetry-allowed angular function --- for the orthorhombic $d$-wave case, one of $k_bk_c$, $k_ak_c$, $k_ak_b$ --- and $\lambda_n(\mathbf{k})$ is left entirely undetermined, save for being invariant under the operations that fix $f_{\rm sym}$. The amplitude is a property of the electronic structure, not of the space group.

The consequence is that symmetry is \emph{permissive}, not \emph{generative}. A magnetic configuration that clears the veto and possesses the requisite spin-exchange operations is \emph{allowed} to show nonrelativistic spin splitting of a definite angular character, with definite enforced nodes. Whether the splitting is large, small, or negligible in a given band and momentum region is decided by $\lambda_n(\mathbf{k})$, which symmetry does not touch. Two materials in the same altermagnetic symmetry class, or two bands of the same material, can therefore share the same enforced nodal surfaces while differing in magnitude by orders of magnitude.

\subsection{Where the magnitude comes from}
\label{subsec:generative_arrangement}
The single-atom anisotropy and its origin in the distortion of the ligand cage are described in Sec.~\ref{sec:introduction}. It is itself independent of symmetry, a fact about the shape of an orbital and not about which operations relate two sites. It does not by itself split a spin. It supplies a shape to be placed, and the non-isotropic, relative placement of that shape between the two sites triggers the splitting from it.

This relative placement is, at the same time, what fixes which operations relate the two sublattices (Sec.~\ref{sec:discussion}), and it is tempting to conclude from this that the crystal symmetry it produces is, after all, what generates the splitting. The following test shows why this is backwards. Suppose the single-atom orbital carries no anisotropy at all, because the ligand cage is undistorted. Then no placement of these isotropic orbitals --- however asymmetric the resulting positions, and however low the resulting crystal symmetry --- can produce a splitting: an isotropic shape looks identical after any rotation, so no placement of it can make $E_{n\uparrow}(\mathbf{k})$ differ from $E_{n\downarrow}(\mathbf{k})$. Lowering the symmetry in this case organizes nothing, because there is nothing there to organize. The relative placement can only trigger a splitting that the orbital's own anisotropy has already made possible; it cannot manufacture one out of orbitals that have no direction to place differently. Symmetry, and the placement that sets it, therefore remain confined to their organizing role even when, as is generic, the arrangement as a whole --- orbital anisotropy together with its relative placement --- is also what generates the splitting's magnitude: the deeper source is the anisotropy of the single-atom orbital, without which no placement and no symmetry, however constructed, could produce a splitting at all. (This complements the formal argument of Sec.~\ref{subsec:Sym-Org}: there, symmetry alone cannot generate a splitting between sublattices that are already electronically distinct; here, no placement or symmetry can generate one when the sublattices carry no orbital anisotropy to begin with.)

In terms of the amplitude $\lambda_n(\mathbf{k})$ of Eq.~\eqref{eq:form_amplitude}, the arrangement --- the tilting and rotation pattern of the surrounding ligands, and the resulting difference in hopping pathways connecting same-spin and opposite-spin sites --- is what makes $E_{n\uparrow}(\mathbf{k})$ and $E_{n\downarrow}(\mathbf{k})$ differ at a generic $\mathbf{k}$ once symmetry allows them to. The magnitude of the splitting scales with the anisotropy of the hopping network,
\begin{equation}
    \lambda_n(\mathbf{k})
    \sim
    \frac{\delta t}{t}\,\times\,(\text{crystal-field / correlation scale}),
    \label{eq:lambda_scaling}
\end{equation}
where $\delta t$ measures the difference between the relevant same-spin and opposite-spin hopping amplitudes.  In perovskites this anisotropy is supplied primarily by the cooperative octahedral distortion; in rutiles by the orthogonal orientation of neighbouring octahedra.  A minimal model built on exactly this mechanism --- an exchange-driven spin polarization combined with a non-magnetic-ion-driven hopping anisotropy $\delta t$, both required simultaneously for a splitting --- has been validated by first-principles calculations on the rutile altermagnet \ce{MnF2} \cite{Lee2025}, independently confirming that $\delta t$ is what carries the amplitude while the relating symmetry there remains intact throughout.  Removing the arrangement anisotropy --- for instance by restoring an equivalence
that makes the two sublattices' hopping networks identical --- gives $\lambda_n(\mathbf{k})\to0$ even though the symmetry labels are unchanged, provided the equivalence is not itself one of the veto operations.  The minimal monoclinic model of Sec.~\ref{sec:minimal_model} realizes this removal as a controlled negative test.

\subsection{Two separate questions}
\label{subsec:two_questions}

We therefore keep two questions explicitly separate throughout this work:
\begin{equation*}
\boxed{
\begin{aligned}
\text{symmetry (permissive):}\quad&
\text{the allowed angular form and the enforced nodes},\\
\text{arrangement (generative):}\quad&
\text{the amplitude }\lambda_n(\mathbf{k})\text{ of the allowed splitting}.
\end{aligned}}
\label{eq:permissive_generative_box}
\end{equation*}
The two are logically independent, and the second does not require the first. Symmetry can permit a splitting whose amplitude the arrangement leaves negligible; conversely, an anisotropic arrangement can generate a finite splitting with no symmetry to organize it. The clearest test of the generative role is therefore a case in which symmetry supplies nothing at all --- a triclinic ($P1$) crystal, where the point group contains only the identity, yet an anisotropic ligand arrangement still produces spin splitting, which disappears when the arrangement is removed. That case is examined in Sec.~\ref{sec:triclinic}, where it also fixes the base of the hierarchy.

For \ce{LaMnO3}, whose large cooperative Jahn--Teller distortion and robust Mn moment supply a strong generative arrangement, the symmetry form and a substantial magnitude coincide, and the three magnetic orders differ in the \emph{orientation} of the $d$-wave form while sharing a common structural and orbital framework. This is the setting in which the hierarchy can be followed directly, and it is why \ce{LaMnO3} is used as the primary vehicle in the sections that follow.

\section{Admissibility controlled by octahedral tilting}
\label{sec:admissibility}

Sections~\ref{sec:permissive_generative} established that the anisotropic arrangement of orbitals between the sublattices, not the magnetic symmetry label, generates the spin-splitting magnitude.  The same arrangement acts at a more basic level as well: it can decide whether nonrelativistic spin splitting is admissible at all.  This section makes that point with two families, \ce{KV2Se2O} and \ce{KCuF3}, in which changing the magnetic order --- or, at fixed order, changing a structural polytype --- switches altermagnetism on and off entirely.  In every case the switch is the Level-I veto of Sec.~\ref{subsec:level1}: whether a pure translation of the parent structure connects the two opposite-spin sublattices.  What controls the presence of that veto, in turn, is the octahedral tilting pattern of the parent lattice.  These systems therefore complement \ce{LaMnO3}: there the veto is always cleared and the interest lies in reconstruction; here the veto itself is toggled, and the analysis terminates at Level~I.

\subsection{The translation veto}
\label{subsec:translation_veto}

Recall from Sec.~\ref{subsec:level1} that a pure translation $T_{\mathbf t}$ exchanging the spin sublattices, Eq. \eqref{eq:X_translation}, leaves the crystal momentum unchanged and therefore forces same-$\mathbf{k}$ degeneracy, Eq. \eqref{eq:translation_degeneracy}, throughout the Brillouin zone.  Any additional spin-exchanging rotation, glide, or screw the structure may contain is then powerless: it relates $0$ to $-0$ and organizes no finite texture.  Admissibility is thus a veto that must be cleared before the reconstruction of Secs.~\ref{sec:bands}--\ref{sec:enforced_accidental} becomes meaningful.

Whether the veto is active is decided by two independent facts. First, \emph{which} pure translation restores the parent structure --- this is fixed by the parent lattice, in particular by its interlayer stacking and octahedral tilting.  Second, whether that translation is spin exchanging under a given magnetic decoration --- this is fixed by the magnetic order.  The veto applies only when the structurally available translation happens to connect opposite-spin sublattices. The examples below hold one factor fixed and vary the other.

\subsection{The structurally available translation}
\label{subsec:available_translation}
In \ce{KV2Se2O} and the D polytype of \ce{KCuF3} the coordination octahedra are essentially untilted and the interlayer orbital pattern repeats along the stacking direction, so the conventional cell contains only two metal atoms.  C-type order is directly represented within this conventional cell; A- and G-type order, both antiferromagnetic along $c$, instead require a $1\times1\times2$ supercell to accommodate the doubled periodicity.  For consistency we describe all three orders within this doubled cell throughout, although the C-type calculation itself was carried out in the conventional cell.  The parent structure is therefore restored by the simple half-$c$ translation
\begin{equation*}
    \ce{KV2Se2O},\ \text{D-}\ce{KCuF3}:\quad
    T_{(0,0,1/2)} .
    \label{eq:adm_Tc}
\end{equation*}

In the A polytype of \ce{KCuF3} the Jahn--Teller/orbital pattern is rotated from one layer to the next, so $T_{(0,0,1/2)}$ no longer restores the structure.  The parent-restoring translation is instead the body diagonal
\begin{equation*}
    \text{A-}\ce{KCuF3}:\quad
    T_{(1/2,1/2,1/2)} .
    \label{eq:adm_Tbody}
\end{equation*}
These two translations are the only candidate vetoes for the respective structures.  Which magnetic orders they suppress then follows purely from the spin parity each order assigns to the relevant translation.

\subsection{Order- and polytype-selected admissibility}
\label{subsec:selection}

Three collinear orders are relevant.  A-type is ferromagnetic within the layer and antiferromagnetic along $c$; C-type is antiferromagnetic within the layer and ferromagnetic along $c$; G-type is antiferromagnetic in all three directions.  Testing the structurally available translation against each order gives the pattern collected in Table~\ref{tab:kcuf3}.

\begin{table}[tb]
\caption{Altermagnetic admissibility of the imposed collinear orders, set by whether the structurally available parent-restoring translation $T$ is spin exchanging under each order.  \ce{KV2Se2O} and D-\ce{KCuF3} share $T_{(0,0,1/2)}$; A-\ce{KCuF3} has $T_{(1/2,1/2,1/2)}$.}\label{tab:kcuf3}
\begin{tabular}{@{}llccc@{}}
\toprule
System & $T$ & A-type & C-type & G-type \\
\midrule
\ce{KV2Se2O}      & $T_{(0,0,1/2)}$     & non-AM & AM     & non-AM \\
D-\ce{KCuF3}      & $T_{(0,0,1/2)}$     & non-AM & AM     & non-AM \\
A-\ce{KCuF3}      & $T_{(1/2,1/2,1/2)}$ & non-AM & non-AM & AM     \\
\bottomrule
\end{tabular}
\end{table}

\paragraph{A-type: non-altermagnetic in every case.}
A-type order is antiferromagnetic along $c$, so any translation with a half-$c$ component exchanges the two spin sublattices.  In \ce{KV2Se2O} and D-\ce{KCuF3} this is $T_{(0,0,1/2)}$; in A-\ce{KCuF3} it is $T_{(1/2,1/2,1/2)}$, whose half-$c$ component acts the same way.  In all three the available translation is spin exchanging,
\begin{equation*}
    \text{A-type:}\quad
    T:\ \uparrow\leftrightarrow\downarrow
    \ \Longrightarrow\
    \Delta_n(\mathbf{k})=0 ,
    \label{eq:adm_A}
\end{equation*}
so A-type order is vetoed and non-altermagnetic throughout, regardless of polytype.

It is worth being explicit about why the veto takes precedence when both kinds of operation are present, since A-type order in these compounds in fact possesses the rotation that an altermagnet would require.  Viewed through that rotation alone, the two spin sublattices are related by a spin-exchanging screw, and one might expect an altermagnetic splitting to follow.  It does not, because the same decoration also leaves a parent-restoring translation relating the sublattices, and that translation holds the two spin channels degenerate at every momentum.  A rotation relating opposite-spin sublattices can only interchange bands that are already separated in energy; where the translation has kept them degenerate, there is nothing for the rotation to interchange, and no splitting appears. The presence of a sublattice-relating rotation is therefore not sufficient on its own: whether it has any effect depends on whether a translation or inversion has already been removed, which is precisely what the Level-I veto tests and what the octahedral tilting decides. This ordering --- admissibility before organization --- is the reason the hierarchy resolves the veto first. The local anisotropy is not what is missing in these cases. In \ce{KCuF3} every \ce{Cu} sits in a Jahn--Teller-distorted octahedron, yet A-type order gives zero splitting in both polytypes, because the translation relates the two sublattices.

\paragraph{C- and G-type: complementary selection.}
For C- and G-type order the half-$c$ translation determines the outcome, and the two polytypes are opposite.  C-type is ferromagnetic along $c$, so $T_{(0,0,1/2)}$ is spin \emph{preserving}; G-type is antiferromagnetic along $c$, so $T_{(0,0,1/2)}$ is spin \emph{exchanging}.  In \ce{KV2Se2O} and D-\ce{KCuF3}, where $T_{(0,0,1/2)}$ is the available translation, this makes C-type altermagnetic and G-type vetoed:
\begin{equation*}
    \text{\ce{KV2Se2O}, D-\ce{KCuF3}:}\quad
    C=\mathrm{AM},\qquad
    G=\mathrm{non\!-\!AM}.
    \label{eq:adm_D}
\end{equation*}
In A-\ce{KCuF3} the available translation is instead $T_{(1/2,1/2,1/2)}$.  Its half-$c$ component is now spin exchanging for C-type (which alternates spin along the body diagonal through the in-plane checkerboard) and spin preserving for G-type, reversing the selection:
\begin{equation*}
    \text{A-\ce{KCuF3}:}\quad
    C=\mathrm{non\!-\!AM},\qquad
    G=\mathrm{AM}.
    \label{eq:adm_A_poly}
\end{equation*}
The selection is therefore governed entirely by which translation the structure makes available.  \ce{KV2Se2O} and D-\ce{KCuF3} are equivalent in the sense relevant here --- both restored by $T_{(0,0,1/2)}$ --- and obey the same rule.  The rotated stacking of A-\ce{KCuF3} replaces that translation by the body diagonal and thereby inverts the C/G outcome, without any change in chemistry.

The altermagnetism of \ce{KV2Se2O} is established independently by experiment and first-principles calculation \cite{Jiang2025}, and we take it here as a known realization in which the surviving $T_{(0,0,1/2)}$ selects the admissible order. For the \ce{KCuF3} polytypes we are not aware of a corresponding analysis in the literature; the A/D comparison and its reversed C/G selection are established by our own calculations (band structures in SM, Sec.~S7).
 
\subsection{Tilting as the physical origin of admissibility}
\label{subsec:tilting}

The comparison with \ce{LaMnO3} identifies the underlying mechanism.  In \ce{KV2Se2O} and both \ce{KCuF3} polytypes a pure translation restores the parent lattice --- $T_{(0,0,1/2)}$ or $T_{(1/2,1/2,1/2)}$ --- precisely because the octahedra are untilted or only simply stacked.  That surviving translation is what makes admissibility fragile: for each structure there is always a magnetic order that renders it spin exchanging and is therefore vetoed (A-type always, and one of C/G depending on the stacking).

\ce{LaMnO3} is the opposite limit.  Its parent structure carries a strong \mbox{GdFeO$_3$}-type distortion --- the cooperative Jahn--Teller elongation combined with the anisotropic rotation and tilting of the \ce{MnO6} octahedra.  After any candidate translation the tilted octahedra no longer coincide with the parent lattice, so \emph{no} pure translation connects the sublattices at fixed $\mathbf{k}$.  In particular the half-$c$ translation $T_{(0,0,1/2)}$, which vetoes A-type order in the untilted systems, is simply not a symmetry of tilted \ce{LaMnO3}.  The Level-I veto is thus structurally absent for all of the A-, G-, and C-type orders, every one of them clears admissibility, and the analysis passes to the reconstruction of Sec.~\ref{sec:bands}.  This is why \ce{LaMnO3} can display three distinct altermagnetic textures where the untilted systems admit at most one of the C/G pair: tilting has removed the veto that would otherwise suppress two of the three orders.

The unifying statement is that octahedral tilting controls altermagnetic admissibility:
\begin{equation*}
\boxed{
\begin{array}{c}
\text{untilted / simply stacked (\ce{KV2Se2O}, \ce{KCuF3})}
\\
\Rightarrow\text{parent translation survives}
\ \Rightarrow\text{veto available; A-type and one of C/G suppressed}
\\[2mm]
\text{tilted (\ce{LaMnO3}, \ce{GdFeO3} type)}
\\
\Rightarrow\text{parent translation destroyed}
\ \Rightarrow\text{no veto; A, C, G all admissible}.
\end{array}}
\label{eq:tilting_veto}
\end{equation*}
This is the admissibility-level counterpart of the permissive/generative distinction of Sec.~\ref{sec:permissive_generative}.  There the anisotropic arrangement generated the magnitude of an allowed splitting; here the same structural anisotropy --- octahedral tilting --- decides whether the splitting is allowed at all, by destroying the translation that would otherwise restore the two spin sectors at every $\mathbf{k}$. The arrangement thus enters at both levels of the hierarchy: it determines whether the Level-I veto is cleared and it sets the Level-III magnitude.  No microscopic exchange argument is needed for either statement; once a magnetic decoration and a parent structure are specified, admissibility follows from whether tilting has removed the degeneracy-restoring translation.

\section{Illustration in orthorhombic \texorpdfstring{\ce{LaMnO3}}{LaMnO3}}
\label{sec:bands}

Orthorhombic \ce{LaMnO3} realizes the orthorhombic tier of Sec.~\ref{subsec:tiers} in an unusually controlled way. Its nonmagnetic $Pbnm$ structure fixes a single network of spatial operations, while A-, G-, and C-type collinear order assign different spin-sublattice parities to those same operations. Changing the magnetic decoration therefore leaves the crystallographic symmetry unchanged while altering the magnetic space group: it reassigns the fixed operation set between the spin-preserving and spin-exchanging sectors, and thereby reorients the enforced nodal planes and the resulting \textit{d}-wave texture.  This makes \ce{LaMnO3} a direct illustration of the plane-supplier principle and the cyclic $D_{2h}$ algebra derived in Sec.~\ref{sec:symmetry_hierarchy}.

We state at the outset what is and is not new here. The spatial operations interchanging opposite-spin Mn sites, and the resulting spin-split and spin-degenerate momentum directions, were computed for \ce{LaMnO3} and related \ce{LaMO3} perovskites by Okugawa \textit{et al.}\ before the current altermagnetic terminology was established \cite{Okugawa2018}. We take those operation mappings and split-line assignments as established. Our contribution in this section is twofold: first, to organize the three magnetic orders through the cyclic parity algebra of Eq.~\eqref{eq:cyclic_triple}, which fixes the admissible $(p,e,e)$ pattern and identifies the excluded configuration; and second, to reconstruct from the complete spin-exchange sector the pair of enforced nodal planes and the three-dimensional $d$-wave orientation for each order --- a geometrical step not carried out in the earlier work.

\subsection{Computational details}
\label{subsec:methods}

The band structures reported below serve only to verify the symmetry-derived relations; the $p/e$ assignments and nodal planes are determined independently from the real-space magnetic decoration, and the calculated eigenvalues provide an independent check of the symmetry construction rather than the basis from which the symmetry assignments are inferred. All three magnetic configurations share the same underlying \mbox{GdFeO$_3$}-type distorted structure; only the spin decoration of the \ce{Mn} sublattice differs, and the space-group label and lattice setting recognized by the electronic-structure code are reduced accordingly for each decoration (SM, Sec.~S2). Full computational details --- exchange--correlation functional, GGA$+U$ parameters, muffin-tin radii, $k$-point meshes, and convergence criteria, for this and every other calculation reported in this work --- are given in SM, Sec.~S1.

\subsection{The cyclic parity algebra and its admissible patterns}
\label{subsec:cyclic}

In the $Pbnm$ setting the operations that relate the two spin sublattices are, as a point-group idealization, the three orthogonal twofold rotations $C_{2a}$, $C_{2b}$, $C_{2c}$.  This $D_{2h}$ frame is what acts transparently in reciprocal space: each $C_{2\alpha}$ maps $\mathbf{k}$ to $C_{2\alpha}\mathbf{k}$ and, for a spin-exchange assignment, relates $\Delta_n$ at the two momenta with a sign change. The bare $C_{2\alpha}$ are not, however, symmetry operations of the nonsymmorphic space group.  With $a<b<c$ in the $Pbnm$ setting, the actual crystal operations are the three twofold screws
\begin{equation}
    S_a=\{C_{2a}\,|\,(\tfrac12,\tfrac12,0)\},\quad
    S_b=\{C_{2b}\,|\,(\tfrac12,\tfrac12,\tfrac12)\},\quad
    S_c=\{C_{2c}\,|\,(0,0,\tfrac12)\},
\label{eq:LMO_ops}
\end{equation}
each a twofold rotation combined with a fractional translation. Because the spin splitting $\Delta_n(\mathbf{k})=E_{n\uparrow}-E_{n\downarrow}$ is real and gauge invariant, the fractional translations do not enter it: each screw acts on $\Delta_n(\mathbf{k})$ exactly as its $C_{2\alpha}$ point part. The $C_{2a},C_{2b},C_{2c}$ frame is therefore the faithful reciprocal-space representation of the real screws $S_a,S_b,S_c$; the translations enter only as Bloch phases and become relevant at the Brillouin-zone boundary, where they govern band sticking.

The nodal \emph{planes} arise through the plane-supplier mechanism of Sec.~\ref{sec:symmetry_hierarchy}.  Taken alone, a spin-exchange $C_{2\alpha}$ leaves invariant only its rotation axis in momentum space, giving a nodal \emph{line}.  A plane requires composing it with inversion.  Here the inversion is supplied not by the crystal but by the nonrelativistic energy itself, which is inversion even, $E_{n\sigma}(-\mathbf{k})=E_{n\sigma}(\mathbf{k})$, so that $\Delta_n(-\mathbf{k})=\Delta_n(\mathbf{k})$ holds whether or not the crystal is centrosymmetric.  The product $\bar1 \cdot C_{2\alpha}$ is then a $\mathbf{k}$-space mirror $m_\alpha$, whose invariant set is a plane:
\begin{equation*}
    \bar1 \cdot {C_{2a}}=m_a,\qquad
    \bar1 \cdot {C_{2b}}=m_b,\qquad
    \bar1 \cdot {C_{2c}}=m_c,
\label{eq:C2_inversion_plane}
\end{equation*}
with invariant planes $k_a=0$, $k_b=0$, $k_c=0$ (the $bc$, $ca$, $ab$ planes).  For a spin-exchange screw $S_\alpha$ this forces $\Delta_n=-\Delta_n$ on the corresponding plane, hence a symmetry-enforced nodal plane.  It is in this precise sense that multiplication by the reciprocal-space inversion turns the $C_2$ line supplier into a plane supplier.

Since parity is multiplicative, Eq.~\eqref{eq:eta_multiplicative_general}, and the three screws close exactly, $S_c=S_aS_b$, the plane-suppliers obey the cyclic constraint
\begin{equation}
    \eta(S_a)\,\eta(S_b)\,\eta(S_c)=+1 ,
    \qquad
    \eta(S_c)=\eta(S_a)\,\eta(S_b),
    \label{eq:LMO_triple}
\end{equation}
the space-group realization of the $D_{2h}$ cyclic algebra $\eta(C_{2a})\eta(C_{2b})\eta(C_{2c})=+1$ of Sec.~\ref{sec:symmetry_hierarchy}.  The admissible parity assignments of $(S_a,S_b,S_c)$ are therefore the three permutations of $(p,e,e)$; each of the former selects one spin-preserving operation and two spin-exchanging ones.  By the plane-supplier mechanism, each $(p,e,e)$ pattern supplies exactly two enforced nodal planes and hence a genuine four-sector orthorhombic $d$-wave texture, with no accidental surface required.

The three magnetic orders realize the three permutations,
\begin{equation*}
    (p,e,e)_A
    \;\longrightarrow\;
    (e,p,e)_G
    \;\longrightarrow\;
    (e,e,p)_C ,
    \label{eq:PX_cycle}
\end{equation*}
for A-, G-, and C-type order respectively, cyclically moving the single spin-preserving operation.  This cyclic reassignment is the central organizing statement of this section, and the calculations below test it directly.

The cyclic algebra presupposes that all three orders are admissible in the first place, i.e.\ that the Level-I veto of Sec.~\ref{subsec:level1} is cleared for each.  This is a nontrivial property of \ce{LaMnO3} and is a direct consequence of its \mbox{GdFeO$_3$}-type octahedral tilting: the tilting destroys the pure translations that would otherwise connect opposite-spin sublattices at fixed $\mathbf{k}$, so no order is vetoed.  In closely related but untilted perovskite-type structures the same half-$c$ translation survives and does veto certain orders --- suppressing A-type universally and one of the C/G pair depending on the stacking --- so that only a single altermagnetic order remains.  The three distinct altermagnetic textures obtained here for \ce{LaMnO3} are therefore possible precisely because tilting removes that veto; we return to this point, and to the untilted counterexamples \ce{KV2Se2O} and \ce{KCuF3}, in Sec.~\ref{sec:admissibility}.

\subsection{Momentum paths that separate the parent operations}
\label{subsec:band_paths}

To distinguish the actions of the parent operations directly, we use four symmetry-related momenta in the $k_c=1/2$ plane,
\begin{equation*}
\begin{aligned}
    R_1&=\left(\tfrac12,\tfrac12,\tfrac12\right),&
    R_2&=\left(-\tfrac12,\tfrac12,\tfrac12\right),\\
    R_3&=\left(\tfrac12,-\tfrac12,\tfrac12\right),&
    R_4&=\left(-\tfrac12,-\tfrac12,\tfrac12\right),
\end{aligned}
\label{eq:R_points}
\end{equation*}
whose point mappings under the $C_{2\alpha}$ operations are
\begin{equation*}
\begin{array}{*{20}{l}}
  {\bar 1 \cdot {C_{2a}}\left[ {\Gamma  - {R_1}} \right] = \left[ {\Gamma  - {R_3}} \right]}&{{\text{or}}\;{m_a}\left[ {\Gamma  - {R_1}} \right] = \left[ {\Gamma  - {R_3}} \right],} \\ 
  {\bar 1 \cdot {C_{2b}}\left[ {\Gamma  - {R_1}} \right] = \left[ {\Gamma  - {R_2}} \right]}&{{\text{or}}\;{m_b}\left[ {\Gamma  - {R_1}} \right] = \left[ {\Gamma  - {R_2}} \right],} \\ 
  {{C_{2c}}\left[ {\Gamma  - {R_1}} \right] = \left[ {\Gamma  - {R_4}} \right],}&{{C_{2c}}\left[ {\Gamma  - {R_2}} \right] = \left[ {\Gamma  - {R_3}} \right].} \\ 
  {{m_b}\left[ {X - {R_1}} \right] = \left[ {X - {R_3}} \right],}&{{m_a}\left[ {Y - {R_1}} \right] = \left[ {Y - {R_2}} \right]}, 
\end{array}
\label{eq:R_mappings}
\end{equation*}
where $X=\left(\tfrac12,0,0\right)$ and $Y=\left(0,\tfrac12,0\right)$.  Because each operation can be spin preserving or spin exchanging depending on the magnetic order, the same pair of paths changes its spin relation when the order is changed, while the underlying crystallographic operation is exactly the same.  This is the direct band-level test of the parity assignments.

\subsection{A-, G-, and C-type order}
\label{subsec:AGC}

\begin{figure*}[tb]
    \centering
    \includegraphics[width=\textwidth]{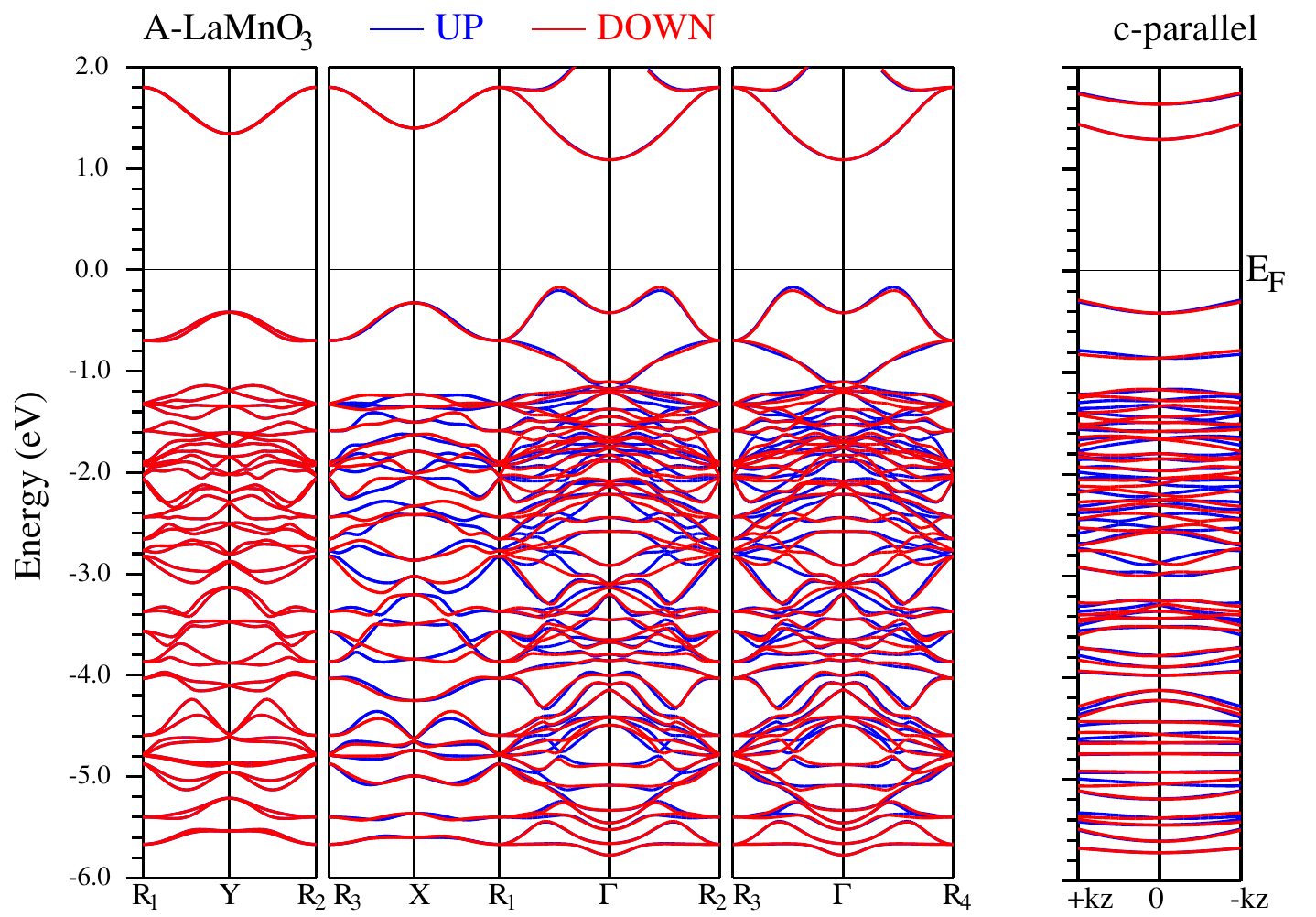}
    \caption{Spin-resolved band structure of A-type \ce{LaMnO3}
    without SOC, following $(S_a,S_b,S_c)=(p,e,e)$: the
    $S_b$- and $S_c$-invariant planes are nodal, while $S_a$
    acts within the same spin sector.}
    \label{fig:A_bands}
\end{figure*}

\paragraph{A-type: $(p,e,e)$.}
Here $\eta_A(S_a)=p$ while $\eta_A(S_b)= e$ and $\eta_A(S_c)= e$.  The enforced nodal planes are therefore the $ca$ and $ab$ planes, and the lowest-order texture is
\begin{equation*}
    \Delta_A(\mathbf{k})\sim k_bk_c .
    \label{eq:Delta_A}
\end{equation*}
The calculated bands (Fig.~\ref{fig:A_bands}) follow this pattern: along $R_1$--$Y$--$R_2$ the two spin channels coincide, as required by $\eta_A(m_a)= p$, whereas along the transverse $m_b$-related path the momenta are related within the opposite spin sector. Because $\eta_A(S_c)= e$, momenta related by $k_c\to-k_c$ are connected by spin exchange and the $ab$ plane is nodal.  The $c$-screw is spin exchanging, consistent with the cyclic relation $\eta_A(S_c)=\eta_A(S_a)\eta_A(S_b)= e$ and with the interchange of the spin-up and spin-down dispersions along the associated directions.

\paragraph{G-type: $(e,p,e)$.}
The crystal is unchanged, but $\eta_G(S_a)= e$ now, the sublattices while $\eta_G(S_b)= p$.  The enforced planes become the $ab$ and $bc$ planes, and
\begin{equation*}
    \Delta_G(\mathbf{k})\sim k_ak_c .
    \label{eq:Delta_G}
\end{equation*}
The calculated bands reverse the transverse behaviour found for A-type order, without any change in crystallographic symmetry --- a direct demonstration that the nodal structure follows the spin-sublattice parity, not the parent operation alone.  Since $S_c= e$ is unchanged between A and G type, the $ab$ plane remains nodal and the $c$-dependent behaviour is common to both orders; the $c$-screw remains spin exchanging, $\eta_G(S_c)= e$.

\begin{figure*}[tb]
    \centering
    \includegraphics[width=\textwidth]{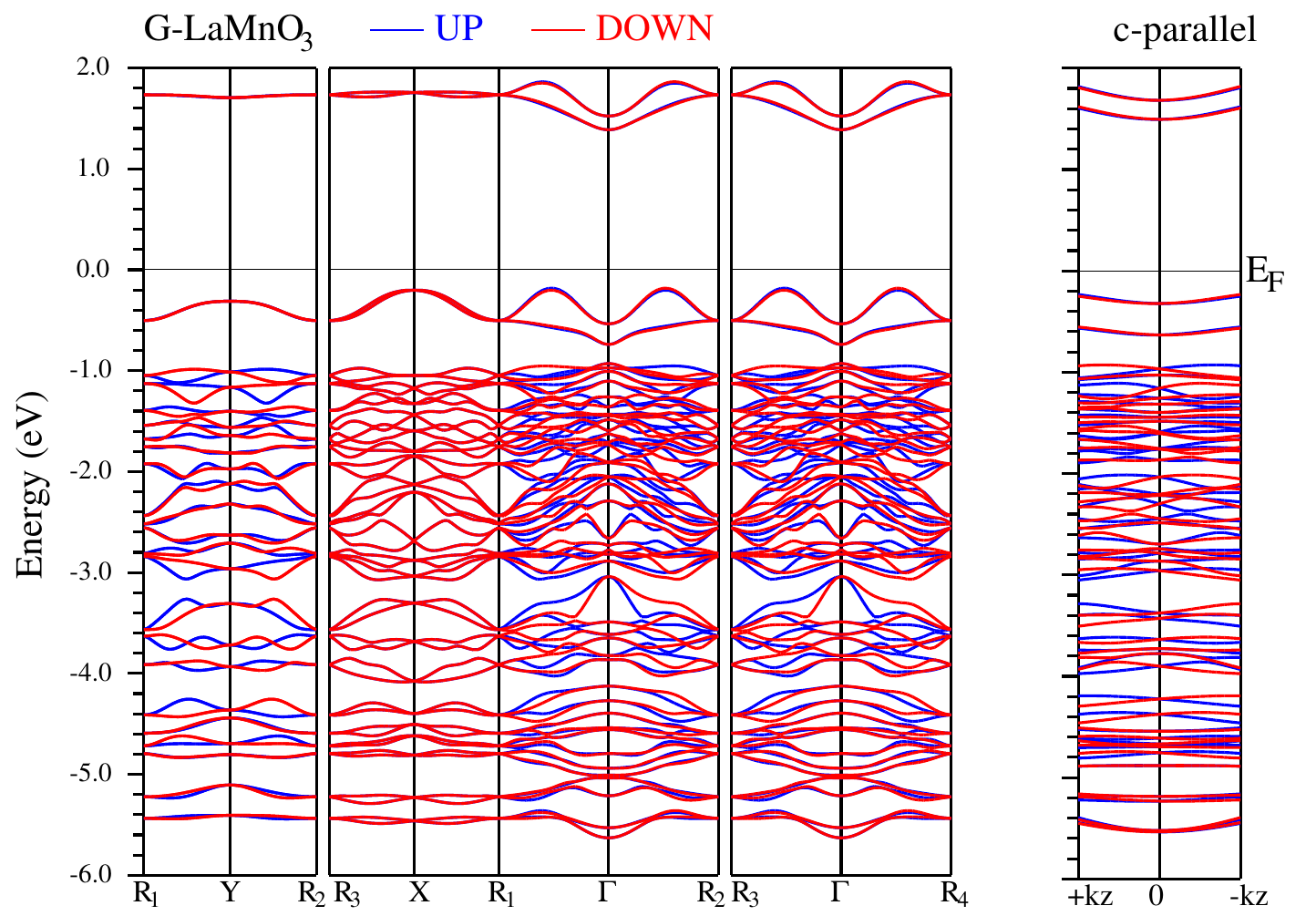}
    \caption{Spin-resolved band structure of G-type \ce{LaMnO3}
    without SOC.  Relative to A-type order the $S_a$ and $S_b$
    parities are exchanged while $S_c$ remains spin exchanging,
    consistent with $(e,p,e)$.}
    \label{fig:G_bands}
\end{figure*}

\paragraph{C-type: $(e,e,p)$.}
Both transverse operations are now spin exchanging, so the $bc$ and $ca$ planes are enforced nodal planes and
\begin{equation*}
    \Delta_C(\mathbf{k})\sim k_ak_b .
    \label{eq:Delta_C}
\end{equation*}
Their intersection divides the relevant section into the four alternating $d$-wave sectors.  The behaviour along $c$ is qualitatively different: $\eta_C(S_c)= p$, so momenta related by $k_c\to-k_c$ are connected within the same spin sector, and the calculated bands show same-spin symmetry along such paths.  The resulting ferromagnet-like appearance of the dispersion along $c$ does \emph{not} indicate a ferromagnetic real-space state; it is the band signature of a spin-preserving spatial operation.  Most importantly, the $c$-screw is now spin \emph{preserving}, as the cyclic relation requires,
\begin{equation*}
    \eta_C(S_c)
    =
    \eta_C(S_a)\,\eta_C(S_b)
    = e\times e
    = p ,
    \label{eq:C_Sc_P}
\end{equation*}
yet the state retains its full $d$-wave altermagnetic texture because $S_a$ and $S_b$ remain independent spin-exchanging operations. C-type \ce{LaMnO3} is thus the clearest illustration in this family of why a single selected screw cannot stand in for the complete spin-exchange sector.

\begin{figure*}[tb]
    \centering
    \includegraphics[width=\textwidth]{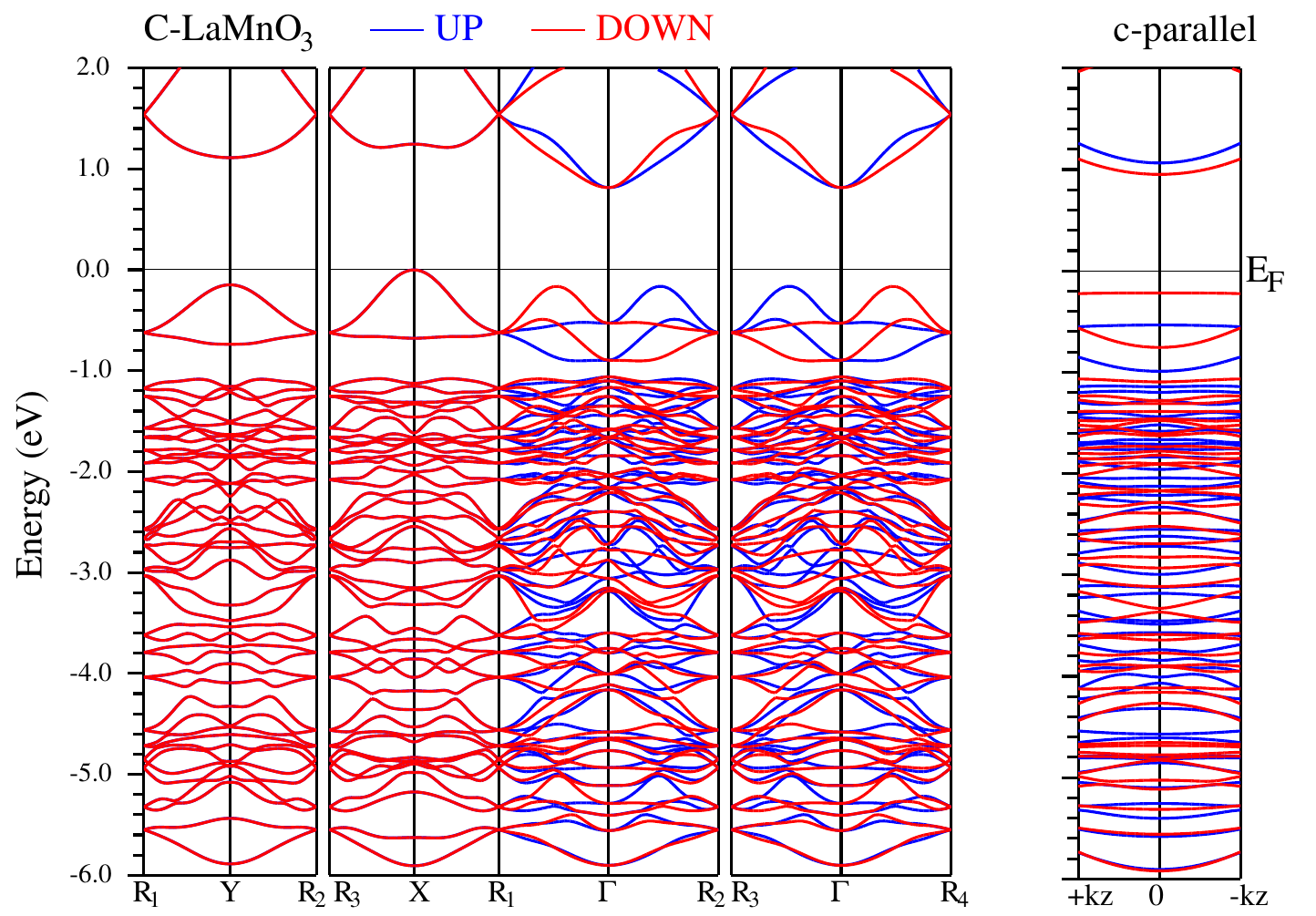}
    \caption{Spin-resolved band structure of C-type \ce{LaMnO3} without SOC, consistent with $(e,e,p)$: $S_a$ and $S_b$ are spin exchanging while $S_c$ is spin preserving, yet the $d$-wave texture survives.  In the last column the splitting keeps a fixed sign, in contrast to the sign reversal seen for A- and G-type.}
    \label{fig:C_bands}
\end{figure*}

One further feature of the calculated bands is worth making explicit. The path $R_1$--$Y$--$R_2$ lies entirely on $k_b=1/2$ and $R_1$--$X$--$R_3$ entirely on $k_a=1/2$ --- the zone-boundary planes of $m_b$ and $m_a$, respectively, invariant by the same mechanism as their zone-centre counterparts (Sec.~\ref{sec:symmetry_hierarchy}). Each path is therefore forced nodal whenever the corresponding mirror is spin exchanging, regardless of which operation genuinely relates its own endpoints: $R_1$--$Y$--$R_2$ is silenced whenever $\eta(m_b)=e$, and $R_1$--$X$--$R_3$ whenever $\eta(m_a)=e$. For A- and G-type order, where $\eta(m_a)$ and $\eta(m_b)$ differ, exactly one path is silenced this way while the other, genuinely exchanged by its own operation, shows the alternating splitting; for C-type order, $\eta_C(m_a)=\eta_C(m_b)=e$, so both paths are silenced and neither shows an altermagnetic signature --- the reason these zone-boundary paths were read, in earlier work, as evidence against altermagnetism in the $(e,e,p)$ pattern. This is a property of these particular paths, not of the texture itself: the enforced planes $bc$ and $ca$ established through Eq.~\eqref{eq:Delta_C}, and the alternating sectors they bound away from these lines, remain exactly as required.

\subsection{Magnetic-order-controlled reorientation of the \texorpdfstring{$d$}{d}-wave texture}
\label{subsec:reorientation}

The three configurations are collected in Table~\ref{tab:AGC}.  Its two sides are obtained independently: the parities from the real-space site mappings and magnetic decorations, the nodal planes and textures from the first-principles eigenvalues.  The nodal planes were therefore predicted from symmetry before the calculated bands were examined, not inferred retrospectively from apparent band crossings.

\begin{table}[tb]
\caption{The three collinear orders of \ce{LaMnO3} as the three $(p,e,e)$ permutations of the cyclic $D_{2h}$ algebra Eq.~\eqref{eq:LMO_triple}, each selecting one spin-preserving axis, two enforced nodal planes, and a lowest-order texture $k_bk_c$, $k_ak_c$, or $k_ak_b$.}\label{tab:AGC}
\begin{tabular}{@{}cccc@{}}
\toprule
Order & $(S_a,S_b,S_c)$ & Nodal planes & $d$-wave plane \\
\midrule
A & $(p,e,e)$ & $ca,\ ab$ & $bc$ \\
G & $(e,p,e)$ & $ab,\ bc$ & $ca$ \\
C & $(e,e,p)$ & $bc,\ ca$ & $ab$ \\
\bottomrule
\end{tabular}
\end{table}

The systematic content is that the three states are not independent realizations built from unrelated mechanisms.  They are three magnetic decorations of one parent crystal structure, differing only in the placement of the single spin-preserving axis.  Cyclically reassigning that axis,
\begin{equation*}
    (p,e,e) \rightarrow (e,p,e) \rightarrow (e,e,p),
\end{equation*}
cyclically moves the pair of enforced nodal planes and reorients the lowest-order texture,
\begin{equation*}
    k_bk_c \;\longrightarrow\; k_ak_c \;\longrightarrow\; k_ak_b .
    \label{eq:texture_cycle}
\end{equation*}
The full construction is thus
\begin{equation*}
\boxed{
\text{magnetic decoration}\rightarrow p/e\text{ reassignment}\rightarrow\text{nodal-plane pair}
\rightarrow d\text{-wave orientation}.
}
\label{eq:lmo_chain}
\end{equation*}

\subsection{Dependence on the \texorpdfstring{$Pnma$/$Pbnm$}{Pnma/Pbnm} axis convention}
\label{subsec:setting_convention}

Because the texture is labelled by crystallographic axes, the choice of setting must be tracked explicitly.  The conventional $Pnma$ and the $Pbnm$ setting used here are alternative descriptions of the same orthorhombic space group with permuted axes, so a layer, chain, or projected motif defined in one setting must be transformed together with the coordinate system.  A geometrically similar projection in a different setting can otherwise represent a different three-dimensional magnetic connectivity, and the $(p,e,e)$ assignment --- and hence the nodal-plane pair --- will appear to change even though the physical state has not.

This setting dependence partly accounts for apparently conflicting reports on the C-type configuration. Yuan and Zunger described the C-type configuration in terms of interconvertible spin-structure motif pairs and obtained vanishing SOC-independent spin splitting \cite{Yuan2023}. In contrast, earlier first-principles calculations  found finite spin-dependent bands for C-type \ce{LaMnO3} \cite{Okugawa2018}; a similar finding was later reported for \ce{YVO3} \cite{Cuono2023}.  In the fixed $Pbnm$ decoration, C-type order is antiferromagnetic within the $ab$ plane and ferromagnetically stacked along $c$; a projection emphasizing the ferromagnetic $c$ chains highlights the single spin-preserving direction ($\eta_C (S_c)= p$) and can make selected local motifs look pairwise interconvertible, but it does not establish equivalence of the complete three-dimensional opposite-spin sublattices.  Once the full crystallographic mappings are retained, $S_a$ and $S_b$ remain spin exchanging and enforce the $bc$ and $ca$ nodal planes, giving finite $\Delta_C\sim k_ak_b$ away from them.  Our calculation reproduces this finite splitting, in agreement with Okugawa \textit{et al.}\ and, we argue, resolving the discrepancy as one of crystallographic setting and projection rather than of physics.  The finite C-type splitting is therefore doubly supported --- by the present calculation and by the earlier independent one --- and is exactly what the $(e,e,p)$ pattern requires.

This is reinforced by a direct band-level signature.  On the path with $k_a,k_b$ fixed and nonzero and $k_c$ running through zero, $\Delta_A\sim k_bk_c$ and $\Delta_G\sim k_ak_c$ are odd in $k_c$ and so reverse sign, while $\Delta_C\sim k_ak_b$ is independent of $k_c$ and keeps a fixed sign --- the sign-reversing versus fixed-sign contrast already visible in the last column of Figs.~\ref{fig:A_bands}--\ref{fig:C_bands} and discussed above.

We note, in addition, that even considered on its own terms the motif-pair method is a symmetry-level admissibility check of the same kind discussed in Sec.~\ref{sec:symmetry_hierarchy}: whether or not its particular verdict on C-type \ce{LaMnO3} is correct, such a method can only rule a texture in or out, and does not by itself address what sets the magnitude of any splitting it permits.

\section{Minimal monoclinic model: \texorpdfstring{$P2$}{P2} and \texorpdfstring{$Pm$}{Pm} \ce{MnO}}
\label{sec:minimal_model}
The orthorhombic tier of \ce{LaMnO3} supplies two enforced nodal planes and therefore hides the distinction between enforced and accidental nodal surfaces: a genuine four-sector $d$-wave texture already follows from symmetry alone.  To expose that distinction, and to test the permissive/generative claim of Sec.~\ref{sec:permissive_generative} directly, we consider the minimal tier of the hierarchy.  By the plane-supplier principle, Eq.~\eqref{eq:plane_supplier}, a monoclinic point group supplies \emph{exactly one} enforced nodal plane, so any second nodal surface a band may show cannot be symmetry-enforced.  Monoclinic symmetry is thus the minimal setting in which enforced and accidental surfaces are forced to coexist, and provides a setting for a negative control.

We therefore introduce two minimal structures, obtained by placing two magnetic \ce{Mn} atoms of opposite spin in a cell whose only nontrivial spatial operation is a twofold rotation ($P2$) or a mirror ($Pm$).  These are not proposed as synthesizable compounds; they are the smallest structures that isolate a single spin-exchange operation together with its ligand environment, so that the role of the surrounding-orbital arrangement can be switched on and off while the magnetic symmetry label is held fixed.

\subsection{Construction of the minimal cells}
\label{subsec:construction}
In the $P2$ realization the two opposite-spin \ce{Mn} sites are related by a twofold rotation $C_2$; in the $Pm$ realization by a mirror $m$.  Each \ce{Mn} is surrounded by an anisotropic ligand (\ce{O}) arrangement that is not shared symmetrically between the two sublattices except through the defining operation. With the ligands present, no pure translation or inversion connects the opposite-spin sublattices, because the anisotropic ligand arrangement removes any such equivalence.  The Level-I veto is therefore cleared, and the operation relating the two spin sites is spin exchanging and moves a generic $\mathbf{k}$,
\begin{equation*}
    \eta(C_2)=-1 \quad(\text{or }\eta(m)=-1), \qquad C_2\mathbf{k}\neq\mathbf{k},
    \label{eq:minimal_X}
\end{equation*}
so the positive-existence condition Eq.~\eqref{eq:positive_existence} is satisfied.  Therefore,  by the monoclinic tier the state is altermagnetic with a single enforced nodal plane --- the invariant plane of $m$, or the plane obtained by promoting the invariant axis of $C_2$ through the even parity of the nonrelativistic energy, as in Eq.~\eqref{eq:plane_supplier}.  Because only one enforced plane is available, any further nodal surface these cells display must be accidental; this is the property exploited in Sec.~\ref{sec:enforced_accidental}.

\subsection{The altermagnetic state}
\label{subsec:minimal_AM}
Both decorated cells develop finite nonrelativistic spin splitting away from the enforced plane, with the two spin channels degenerate on the invariant manifold of the spin-exchange operation and interchanged between momenta related by it,
\begin{equation*}
    E_{n\uparrow}(C_2\mathbf{k})
    =
    E_{n\downarrow}(\mathbf{k}).
    \label{eq:minimal_exchange}
\end{equation*}
The lowest-order allowed form has the single enforced plane as one nodal surface; where a four-sector pattern appears, its second nodal surface is supplied not by symmetry but by the electronic structure. The band-resolved momentum-space maps that make this second surface visible, and that establish its accidental character, are presented in Sec.~\ref{sec:enforced_accidental}.

\subsection{The negative control: stripping the arrangement}
\label{subsec:negative_control}

The generative claim of Sec.~\ref{subsec:generative_arrangement} is that the splitting magnitude is produced by the anisotropic \emph{arrangement} of orbitals between the sublattices, not by the magnetic symmetry label.  The minimal cells allow this to be tested by a controlled deletion.

Remove the ligands and retain only the two \ce{Mn} atoms in their positions.  The bare two-atom cell possesses higher symmetry than the decorated one: with the anisotropic environment gone, an inversion centre (or a pure translation) again relates the two \ce{Mn} sites. In the antiferromagnetic decoration this recovered operation connects the opposite-spin sublattices at fixed $\mathbf{k}$,
\begin{equation*}
    I:\ \uparrow\leftrightarrow\downarrow ,
    \label{eq:minimal_inversion}
\end{equation*}
and, combined with $E_{n\sigma}(-\mathbf{k})=E_{n\sigma}(\mathbf{k})$, enforces same-$\mathbf{k}$ degeneracy throughout the zone.
The bare cell is thus a conventional, fully compensated antiferromagnet with no nonrelativistic spin splitting.

The comparison is controlled because the magnetic order is identical throughout --- two \ce{Mn} atoms of opposite spin --- and only the ligand arrangement is changed.  The passage from the decorated cell to the bare cell is not a change of magnetic configuration but a change of the surrounding-orbital \emph{arrangement}:
\begin{equation*}
\boxed{
\begin{array}{c}
\text{anisotropic arrangement present}
\;\Rightarrow\;
\text{veto cleared},\ \Delta_n\neq0\ \text{(AM)}
\\[1mm]
\Downarrow\ \text{strip arrangement}
\\[1mm]
\text{inversion recovered}
\;\Rightarrow\;
\text{veto active},\ \Delta_n\equiv0\ \text{(conventional AFM)}.
\end{array}}
\label{eq:negative_control}
\end{equation*}

This realizes the permissive/generative distinction as an explicit on/off test.  Stripping the anisotropic arrangement does not merely reduce the magnitude $\lambda_n(\mathbf{k})$ of Eq.~\eqref{eq:lambda_scaling}; by restoring a degeneracy-restoring equivalence it re-activates the Level-I veto and removes the altermagnetism entirely.  The magnetic symmetry label ``two opposite-spin \ce{Mn} atoms related by an operation'' is common to the altermagnetic and the conventional cases; what distinguishes them is solely whether the orbital arrangement supplies a spin-exchange operation with no competing degeneracy-restoring equivalence.  The minimal model thus provides a controlled demonstration that the altermagnetic splitting is generated by the anisotropic arrangement and only shaped --- through its nodes --- by symmetry.

\section{Enforced versus accidental nodal surfaces}
\label{sec:enforced_accidental}

The hierarchy of Sec.~\ref{sec:symmetry_hierarchy} distinguishes two kinds of zero of $\Delta_n(\mathbf{k})$: nodes forced by a spin-exchange operation through Eq.~\eqref{eq:X_enforced_node}, whose existence and location are fixed by symmetry, and accidental nodes that arise from the detailed electronic structure and are constrained only to occur in a symmetry-compatible pattern.  This distinction is usually stated but rarely demonstrated, because in high-symmetry altermagnets every nodal surface a band shows is enforced and there is nothing accidental to exhibit.  The minimal monoclinic model of Sec.~\ref{sec:minimal_model} is constructed to separate the two cases: it supplies exactly one enforced plane, so any further nodal surface must be accidental, and the two kinds can be separated directly.  This section presents that separation as the central empirical result of the work.

We use the $P2$-\ce{MnO} altermagnet and map $\Delta_n(\mathbf{k})$ plane by plane, sweeping the momentum $k_b$ transverse to the invariant plane of the spin-exchange operation (The underlying band structures of the $P2$- and $Pm$-\ce{MnO} models are given in SM, Sec.~S3.). The spin-exchange operation has the $k_b=0$ plane as its invariant manifold, so this is the candidate enforced plane and $b$ is the transverse direction; each map in Fig.~\ref{fig:heatmap} is a constant-$k_b$ section through the $ac$ plane.  Two energy-adjacent bands, labelled $18$ and $19$, are compared throughout.

\subsection{The enforced plane: band-universal}
\label{subsec:enforced_universal}
Equation~\eqref{eq:X_enforced_node} requires the splitting to vanish on the invariant plane of the spin-exchange operation,
\begin{equation*}
    \Delta_n(k_a,0,k_c)=0
    \label{eq:kb0_node}
\end{equation*}
for \emph{every} band $n$, independently of the band index, the dispersion, or the hopping details. By the same mechanism (Sec.~\ref{sec:symmetry_hierarchy}), the zone-boundary plane $k_b=\pm 1/2$ is an equally invariant, parallel copy of this plane and is equally enforced,
\begin{equation*}
    \Delta_n(k_a,\tfrac12,k_c)=0 ,
    \label{eq:kbhalf_node}
\end{equation*}
not a second, independent condition but the same one asked at the zone boundary instead of the zone centre. The maps confirm both directly: $\Delta_n$ vanishes identically across the entire $ac$ plane at $k_b=0$ and, equally, at $k_b=\pm 1/2$, for both band $18$ and band $19$ [Fig.~\ref{fig:heatmap}, leftmost and rightmost panels of each row]. Band-universality is the signature of an enforced node --- symmetry acts on all bands at once, so an enforced surface occupies the same location in every band's map. Both boundary planes behave exactly this way: blank for both bands and, we verified, for every other band inspected.

\subsection{The accidental surface: band-specific in location}
\label{subsec:accidental_location}

\begin{figure*}[tb]
    \centering
    \begin{subfigure}{\textwidth}
        \centering
        \includegraphics[width=\textwidth]{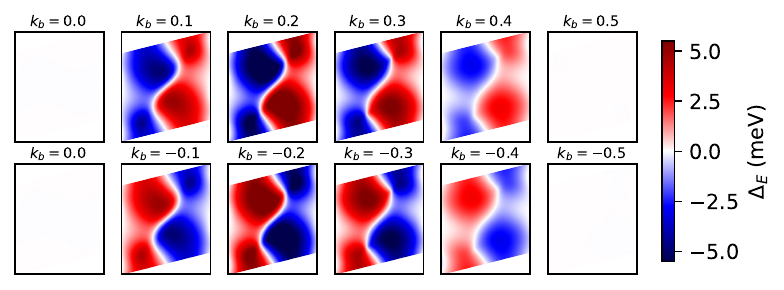}
        \caption{Band $18$.}
        \label{fig:heatmap_b18}
    \end{subfigure}\\[1ex]
    \begin{subfigure}{\textwidth}
        \centering
        \includegraphics[width=\textwidth]{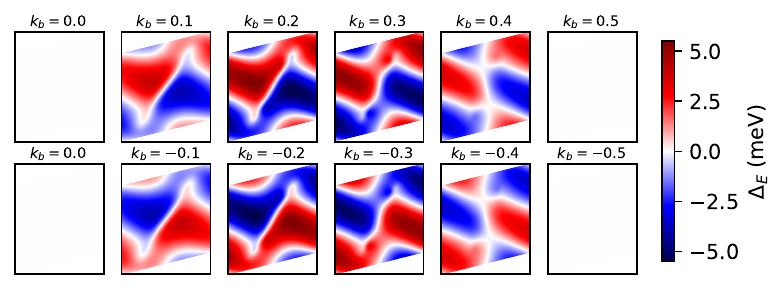}
        \caption{Band $19$.}
        \label{fig:heatmap_b19}
    \end{subfigure}
    \caption{Momentum-space maps of the spin splitting $\Delta_n(\mathbf{k})$ in $P2$-\ce{MnO} without SOC, as constant-$k_b$ sections through the $ac$ plane for two energy-adjacent bands, with $k_b$ from $0$ to $\pm0.5$.  At $k_b=0$ (leftmost panels) and $k_b=\pm0.5$ (rightmost panels) $\Delta_n=0$ identically for both bands --- the enforced, band-universal node.  Away from $k_b=0$ an accidental line appears and migrates, coherent for band $18$ (a) but fragmenting into extra lobes for band $19$ (b).  The sign reversal between $+k_b$ and $-k_b$ identifies a $d$-wave texture.}
    \label{fig:heatmap}
\end{figure*}

Away from $k_b=0$ the sections develop a line along which $\Delta_n$ passes through zero.  This line is not fixed by the monoclinic symmetry, which supplies only the single plane Eq.~\eqref{eq:kb0_node}, and as $|k_b|$ increases the line \emph{migrates} across the $ac$ plane.  Its geometry differs between the two bands: for band $18$ the migrating node forms a relatively simple curve traversing the section, whereas for band $19$ at the same $k_b$ it develops a more structured shape with additional sign reversals [Fig.~\ref{fig:heatmap}].

This has a direct consequence.  Symmetry cannot distinguish band $18$ from band $19$: both transform identically under the spin-exchange operation, so any symmetry-enforced surface would sit at the same momenta in both.  A nodal surface whose location differs between two symmetry-equivalent bands therefore cannot be symmetry-enforced.  The $k_b\neq0$ line is \emph{accidental in location} --- its position is set by the band-specific dispersion, not by the space group.

\subsection{The accidental surface: band-specific in existence}
\label{subsec:accidental_existence}

The band dependence goes further than geometry.  In a given transverse section the accidental line is well developed for one band and weak or fragmented for the other: at intermediate $k_b$ the band $18$ map carries a single clear zero-crossing curve, while the band $19$ map in the same section shows the node broken into disconnected segments or replaced by additional lobes rather than one coherent surface [compare the two rows of Fig.~\ref{fig:heatmap}].  The surface is thus not merely displaced between bands; as a single connected nodal object it is not robustly present in both.

This is a stronger statement than band-specific location.  An enforced node cannot be absent, or reorganize into a different connectivity, for a subset of bands --- Eq.~\eqref{eq:X_enforced_node} applies identically to all of them --- so a surface whose very existence and connectivity depend on the band is accidental \emph{in existence}, not only in location.  The $k_b\neq0$ surface of $P2$-\ce{MnO} is accidental in both senses at once: where it appears, its position is band-specific, and whether it appears as a coherent surface at all is band-specific.

\subsection{The texture is \texorpdfstring{$d$}{d}-wave, not \texorpdfstring{$p$}{p}-wave}
\label{subsec:dwave_not_pwave}

A single nodal line separating a positive from a negative region might be read as a $p$-wave (dipolar) texture.  The maps rule this out. Sections at opposite transverse momenta carry opposite overall sign,
\begin{equation}
    \operatorname{sgn}\Delta_n(k_a,+|k_b|,k_c)
    =
    -\operatorname{sgn}\Delta_n(k_a,-|k_b|,k_c),
    \label{eq:sign_reversal}
\end{equation}
visible in Fig.~\ref{fig:heatmap} as the interchange of the red and blue regions between the $+k_b$ and $-k_b$ rows of panels.  Combined with the sign structure within each section, the full pattern is the four-lobed alternation $+,-,+,-$: an even, $d$-wave texture whose nodal surfaces are the enforced $k_b=0$ plane and the accidental migrating line.  The reversal Eq.~\eqref{eq:sign_reversal} follows from the even parity of the nonrelativistic energy together with the spin-exchange relation, and is exactly what separates the compensated $d$-wave altermagnet from a dipolar $p$-wave state.  The presence of an accidental line therefore does not lower the angular character of the texture.

\subsection{Interpretation and relation to ``continuity-enforced'' nodes}
\label{subsec:continuity_discussion}

The two surfaces of $P2$-\ce{MnO} therefore have qualitatively different origins.  The $k_b=0$ plane is enforced: band-universal, fixed in location, present in every band.  The $k_b\neq0$ line is accidental: band-specific in geometry and band-specific in existence. The minimal monoclinic construction is what makes this contrast visible, since a single enforced plane leaves the second surface unprotected and free to reveal its electronic-structure origin.

This bears on how such secondary nodes are described in the recent literature.  In some analyses --- for instance of \ce{BiFeO3}\cite{Urru2025} --- a second nodal surface that is not enforced by an isolated operation is nonetheless argued to be required by continuity of the band structure and is termed \emph{continuity-enforced}.  Our evidence leads us to a different reading for the present case.  A node that is enforced, even by continuity, cannot simply fail to exist, or change its connectivity, for some bands while persisting for symmetry-equivalent neighbours; yet exactly this band-specific \emph{existence} is what the band $18$/band $19$ comparison shows.  We therefore describe the $k_b\neq0$ surface as accidental in both location and existence, rather than as continuity-enforced, and adopt this terminology throughout.  We state the boundary of the claim in one sentence: this is a considered divergence in terminology grounded in the band-specific existence evidence for $P2$-\ce{MnO}, and it is not intended to deny that continuity can constrain the accidental nodes into a symmetry-compatible pattern once they exist.

There is a further difficulty with taking continuity as a source of enforcement.  If continuity is what forces a surface into existence, one may ask why it forces this surface and not others: continuity by itself fixes no location, and what selects where the node falls is the electronic structure, exactly as for any accidental surface.  A genuinely enforced node does not have this ambiguity --- the $k_b=0$ plane is fixed by the spin-exchange operation at each momentum, band-universally and independently of the dispersion --- and it is precisely this property that the $k_b\neq0$ surface lacks.  Admitting continuity as a form of enforcement therefore blurs the boundary between the two kinds of node rather than drawing it, since the same reasoning would have to grant enforced status to any accidental surface that continuity shapes into a symmetry-compatible pattern.

\section{Triclinic \texorpdfstring{$P1$}{P1}-MnO: accidental nodal surfaces without symmetry}
\label{sec:triclinic}

The analysis so far treats altermagnetism as a question to be decided by symmetry: clear the admissibility veto, find the spin-exchange operations, read off the enforced nodal planes. Applied down to the monoclinic class, this account is complete and self-consistent, and the triclinic class is conventionally left out of it. The omission is not wrong --- with no rotation or mirror to relate the sublattices, a triclinic crystal cannot be an altermagnet --- but it is incomplete, and the incompleteness is instructive. Being unable to be an altermagnet is not the same as having no spin splitting. In setting the triclinic case aside as symmetry-free, one quietly assumes the second from the first. The assumption fails, and a precondition for altermagnetism that the symmetry account never names is what fails with it.

The precondition surfaces first as a question about the accidental surface of the monoclinic tier. A monoclinic altermagnet supplies one enforced plane and requires a second, accidental surface for a four-sector $d$-wave texture. This second surface is usually attributed to the electronic structure and left there. The attribution is correct but its causal order, as commonly stated, is inverted. One says: a monoclinic altermagnet has a single enforced plane, so \emph{in order to} form a four-sector texture a second surface must appear. Read as cause and effect this is backwards, because it makes the altermagnetic texture call into being the surface it requires. The physical order is the reverse. A $p$-wave (two-sector) texture is forbidden for collinear nonrelativistic splitting, since $\Delta_n(-\mathbf{k})=\Delta_n(\mathbf{k})$ excludes an odd-parity form; any splitting a monoclinic crystal develops must therefore be even, and a second surface must already be present for that even texture to exist at all. The second surface is not summoned by the $d$-wave requirement; it must be supplied first, by something other than symmetry, and only then can the single enforced plane organize the result into a $d$-wave altermagnet. If that something is absent, the monoclinic crystal is not a $d$-wave altermagnet: it is a fully compensated ferrimagnet, its accidental surfaces present without a symmetry to organize them into a definite wave character. This is a distinct outcome from what happens when the crystal's one available operation is instead a veto rather than an organizer --- as when the inversion of a centrosymmetric $2/m$ point group is spin exchanging (Sec.~\ref{sec:symmetry_hierarchy}). There $\Delta_n(\mathbf{k})\equiv0$ identically, a conventional antiferromagnet with no surface, enforced or accidental, to be found at all: the veto forecloses not only the enforced plane but the very possibility of a splitting for any accidental surface to occupy.

The triclinic case isolates this ``something'' with nothing else present. In the point group $1$ the only spatial operation is the identity: no rotation, no mirror, no glide, no screw, and no inversion. By the plane-supplier principle, Eq.~\eqref{eq:plane_supplier}, the number of enforced nodal planes is zero. Any nodal surface a band develops is therefore accidental in the strongest sense --- not enforced in location, and not enforced in existence, because no operation constrains it at all. If spin splitting nonetheless occurs in a $P1$ crystal, and if it is removed by removing the ligand arrangement while the $P1$ label is held fixed, then the arrangement, not symmetry, is its origin. This is the conclusion the following calculation establishes.

A version of this same tension has already been noticed directly, though not pursued.  A thinned $(110)$ film of \ce{RuO2} with collinear antiferromagnetic order fails the altermagnetic symmetry criterion --- the translational symmetry the criterion requires is broken by the reduction to a single unit cell --- yet a minimal hopping model built on the microscopic mechanism of Sec.~\ref{subsec:generative_arrangement} still predicts a finite splitting there, a discrepancy left open for future work \cite{Lee2025}.  The triclinic ferrimagnet examined here is the case in which that discrepancy is sharpest and is resolved rather than left open: with no symmetry present at all, to be vetoed or to organize, the arrangement is what remains to answer the question, and the calculation below shows directly that it does.

\subsection{A triclinic $P1$ MnO and its $P\bar1$ counterpart}
\label{subsec:P1_construction}
We consider a triclinic cell containing two inequivalent \ce{Mn} sites and two \ce{O} ligands, with lattice parameters and internal coordinates carrying no special relations, so that the space group is $P1$: symmetry finding returns the identity as the only operation. The two \ce{Mn} sites are thus crystallographically inequivalent, and no translation, inversion, rotation, or mirror relates the two spin sublattices. Alongside this structure we consider a counterpart in which the same two \ce{O} atoms --- unchanged in number and species --- are repositioned to restore an inversion centre relating the two \ce{Mn} sites, so that the space group becomes centrosymmetric, $P\bar1$. The two calculations share the same collinear two-\ce{Mn} magnetic decoration; they differ only in the ligand arrangement.

In the ligand-decorated cell the calculated sublattice moments are unequal, \(4.51\) and \(-4.49\,\mu_B\) on the two \(\ce{Mn}\), with induced \(-1.01\) and \(+0.94\,\mu_B\) on the two \(\ce{O}\). The inequality of the \(\ce{Mn}\) magnitudes is the direct signature that no symmetry relates the sublattices: a symmetry-related pair would be exactly equal in magnitude. The net cell moment is small but finite, \(-0.07\,\mu_B\), as expected for a \(P1\) structure in which no operation enforces compensation. Two points matter for what follows. First, each \(\ce{Mn}\) carries a large, well-developed moment of about \(4.5\,\mu_B\); the small net moment results from the near cancellation of large but unequal sublattice contributions, not from weak magnetic order. The finite residual moment is therefore significant: it confirms that the state is genuinely ferrimagnetic rather than an antiferromagnet whose compensation is enforced by symmetry. Second, the state falls outside the three standard collinear classes: it is not a conventional antiferromagnet, whose opposite-spin sublattices are symmetry-equivalent; not an altermagnet, whose sublattices are related by a rotation or mirror; and not a ferromagnet, whose moments are aligned rather than nearly cancelling. It is thus a nearly compensated ferrimagnet, continuously connected to the fully compensated ferrimagnetic limit discussed recently \cite{Guo2025,Dong2025}, in which the opposite-spin sublattices are connected by no crystal symmetry and compensation, when exact, is not symmetry enforced.

\subsection{Splitting without symmetry, and its collapse under a restored inversion}
\label{subsec:P1_bands}
\begin{figure*}[tb]
    \centering
    \begin{subfigure}{0.49\textwidth}
        \centering
        \includegraphics[width=\textwidth]{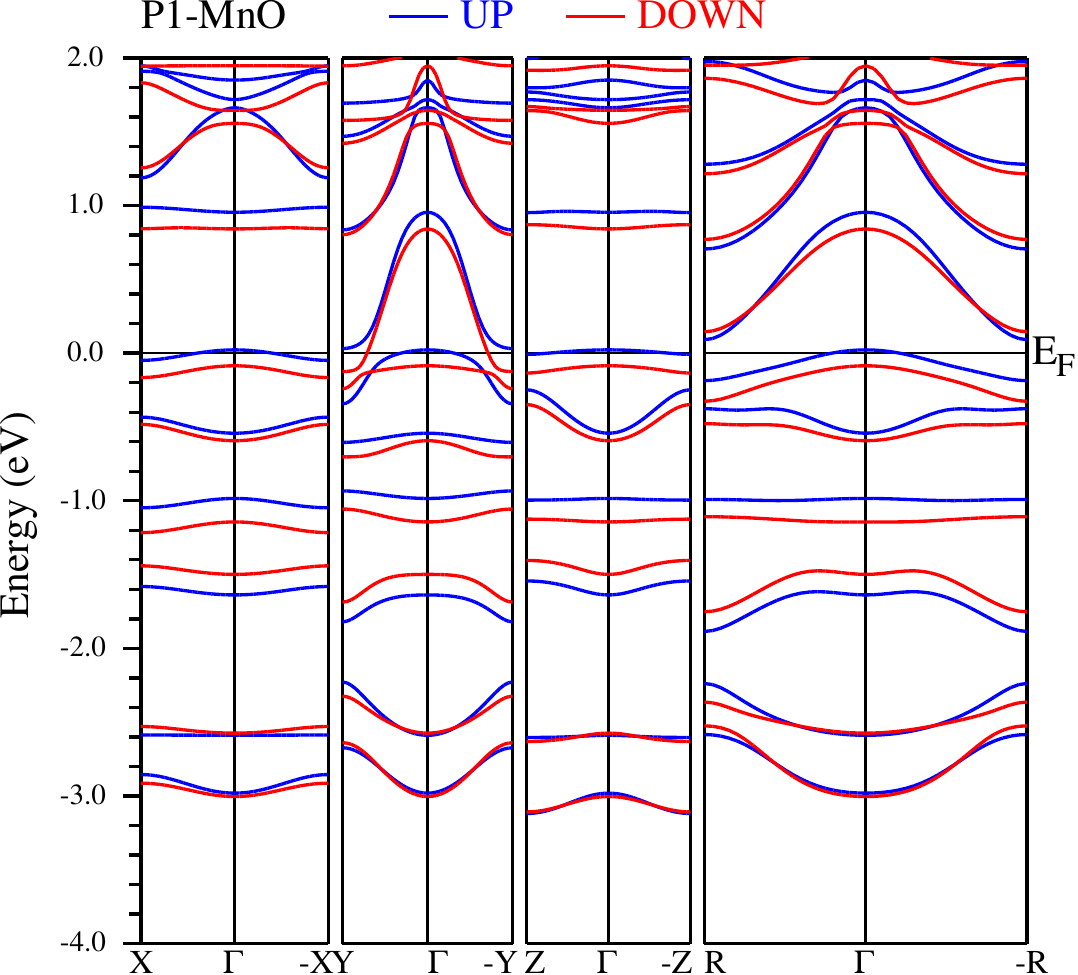}
        \caption{}
        \label{fig:p1_mno}
    \end{subfigure}
    \hfill
    \begin{subfigure}{0.49\textwidth}
        \centering
        \includegraphics[width=\textwidth]{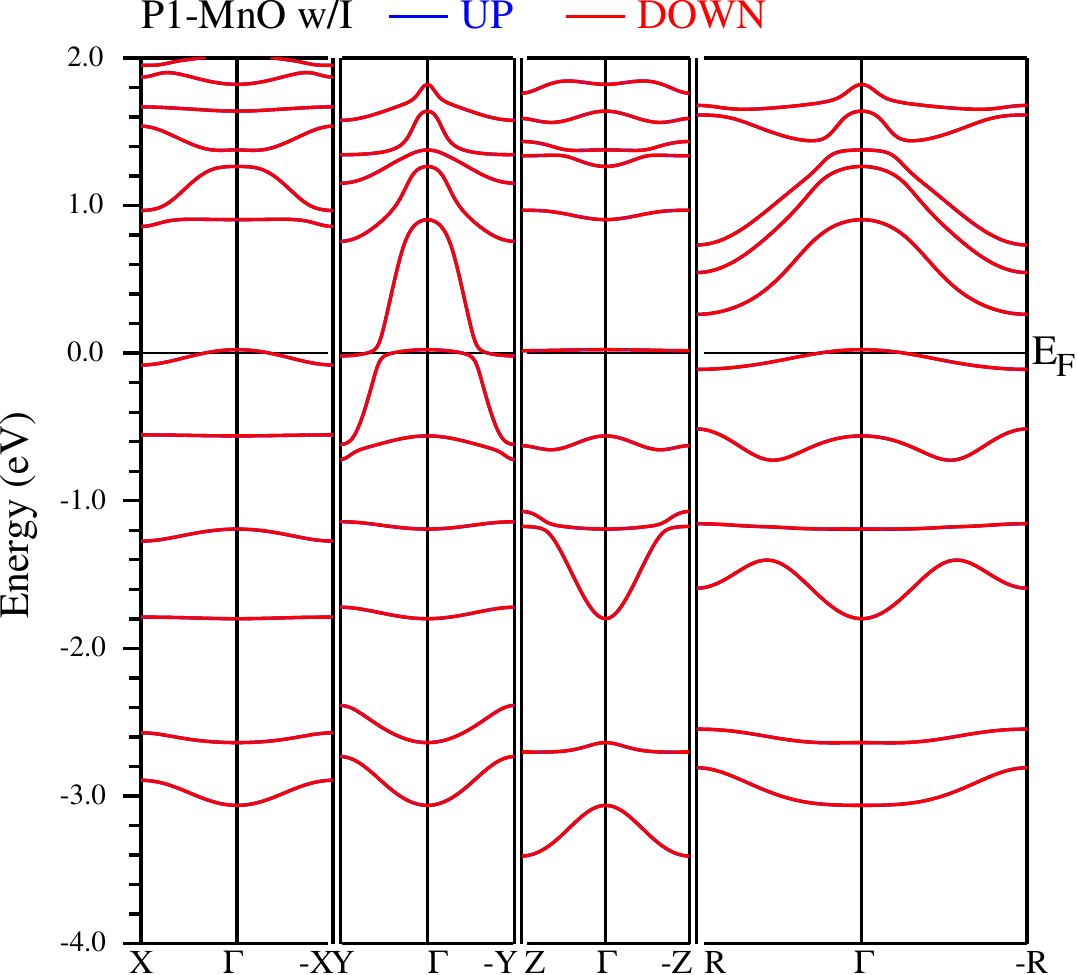}
        \caption{}
        \label{fig:p1_mn}
    \end{subfigure}
    \caption{Spin-resolved band structure of triclinic \ce{MnO} without SOC. (a) With the \ce{O} ligands in their original positions ($P1$), the bands are spin split throughout the zone, with the sign alternating mainly between bands and reversing within only a few of them. (b) With the same ligands repositioned, atom for atom, to restore an inversion centre relating the two \ce{Mn} sites ($P\bar1$), the spin channels coincide, $\Delta_n(\mathbf{k})\simeq 0$, visible as the red dispersion curves overlying the blue ones throughout the path.}
    \label{fig:p1}
\end{figure*}

Figure~\ref{fig:p1} shows the spin-resolved bands of the two structures introduced above. The two calculations share the same collinear two-Mn magnetic decoration; crystallographically, the original structure is \(P1\), whereas repositioning the ligands restores \(P\bar1\). They differ only in the ligand arrangement. With the ligands in their original positions [Fig.~\ref{fig:p1}(a)], the bands are spin split throughout the zone even though the space group contains only the identity. The splitting has a further feature that distinguishes it from a simple ferromagnetic exchange shift and from the symmetry-organized sign alternation of an altermagnet: its dominant sign alternation occurs between bands. Reading up the spectrum, some bands place their spin-down component below spin-up and others place spin-up below; many bands retain a fixed sign, while selected bands also undergo momentum-dependent sign reversals through accidental nodal surfaces. The bands connect as smooth curves, without the derivative kinks that mark a crossing of orbital character, so the energy-ordered index coincides with orbital character along these paths and the spin comparison is a genuine same-orbital splitting. Unlike in a conventional antiferromagnet, where compensation occurs at each \(k\), or in an altermagnet, where symmetry guarantees compensation within each symmetry-matched band after Brillouin-zone integration, no such band-resolved cancellation is enforced in the $P1$ ferrimagnet. Its small net moment emerges only after the spin-resolved occupations are summed over the full set of bands and momenta. The alternating sign of the band splittings is consistent with compensation occurring only after contributions from different bands are combined, rather than with a uniform ferromagnetic exchange shift --- a distribution set by which \ce{Mn} sublattice dominates each band (the site- and orbital-resolved character underlying this distribution is shown, for the $P\bar1$ endpoint, in SM, Sec.~S5), and therefore by the anisotropic arrangement of the magnetic orbitals.

\begin{figure*}[tbh]
    \centering
    \begin{subfigure}{0.49\textwidth}
        \centering
        \includegraphics[width=\textwidth]{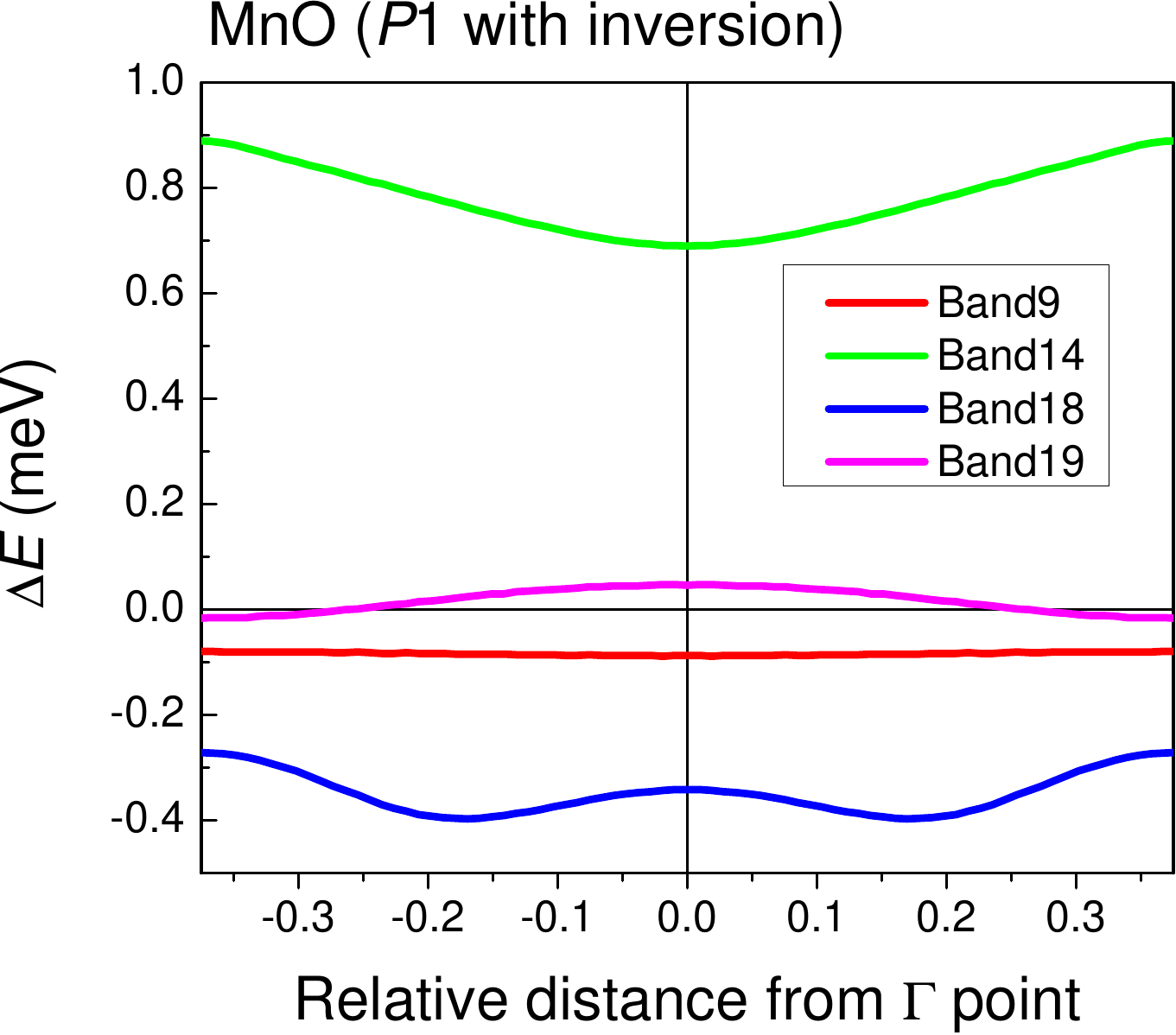}
        \caption{}
        \label{fig:de_inversion}
    \end{subfigure}
    \hfill
    \begin{subfigure}{0.49\textwidth}
        \centering
        \includegraphics[width=\textwidth]{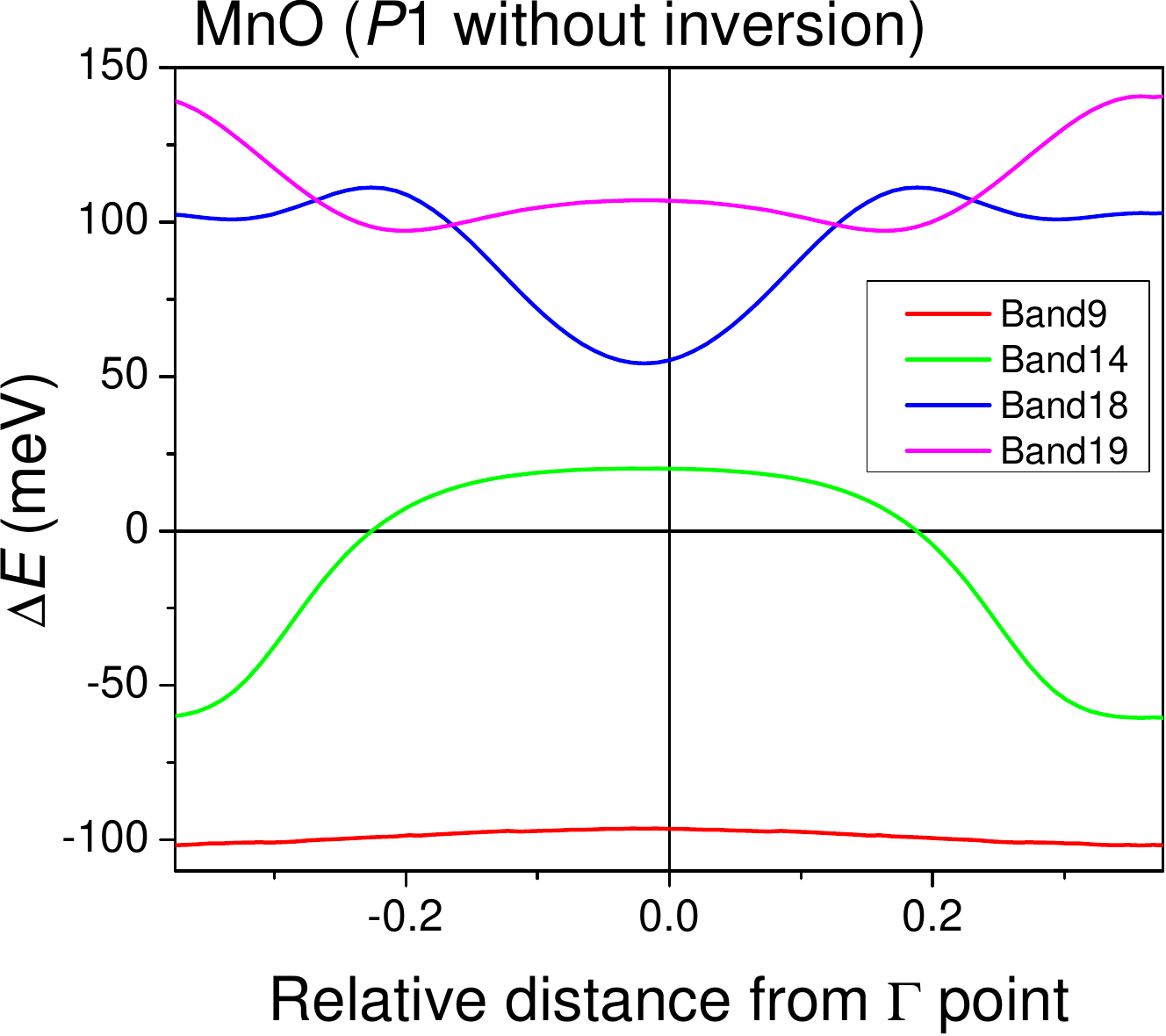}
        \caption{}
        \label{fig:de_p1}
    \end{subfigure}\\[1ex]
    \begin{subfigure}{0.49\textwidth}
        \centering
        \includegraphics[width=\textwidth]{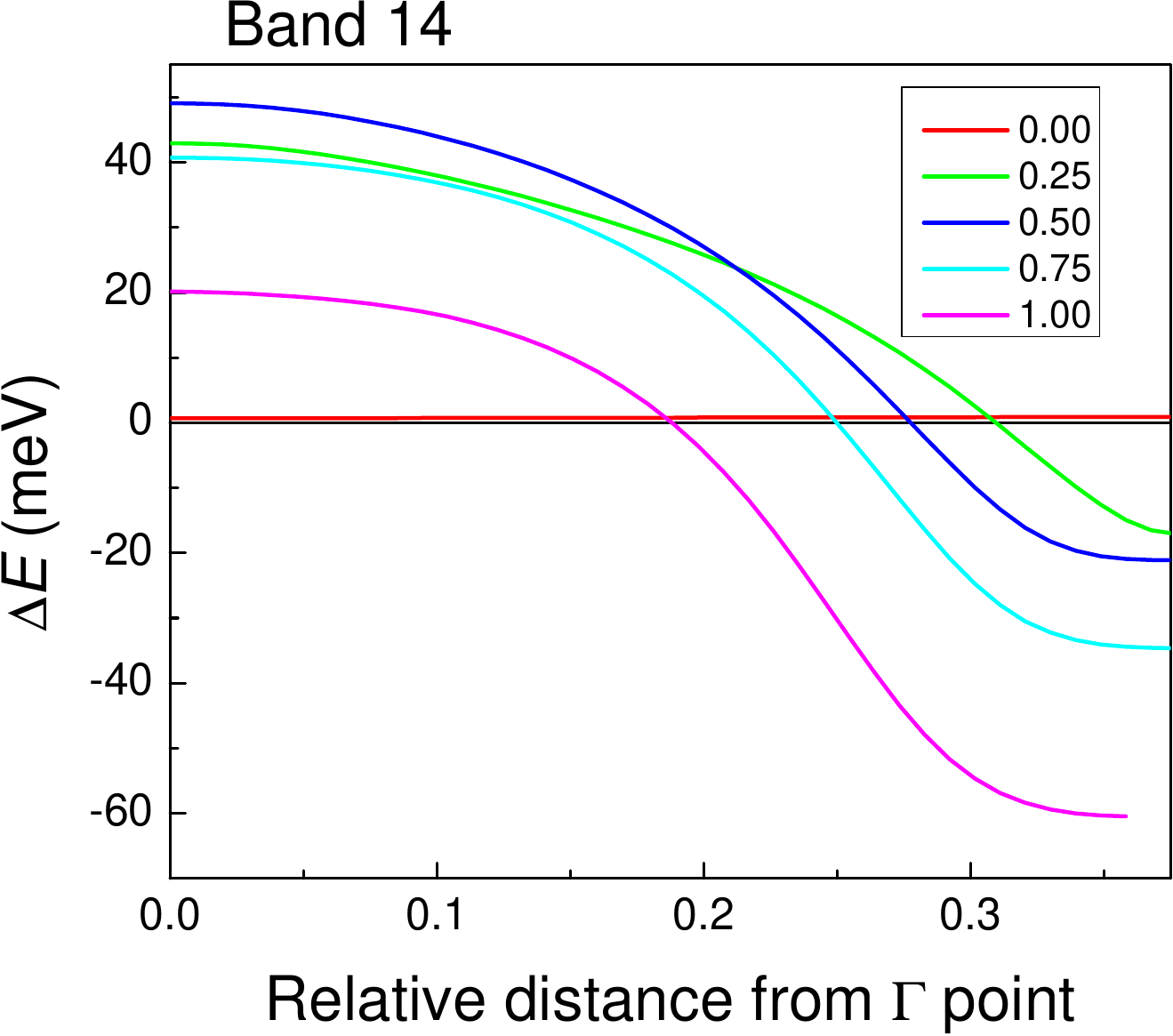}
        \caption{}
        \label{fig:de_band14}
    \end{subfigure}
    \hfill
    \begin{subfigure}{0.49\textwidth}
        \centering
        \includegraphics[width=\textwidth]{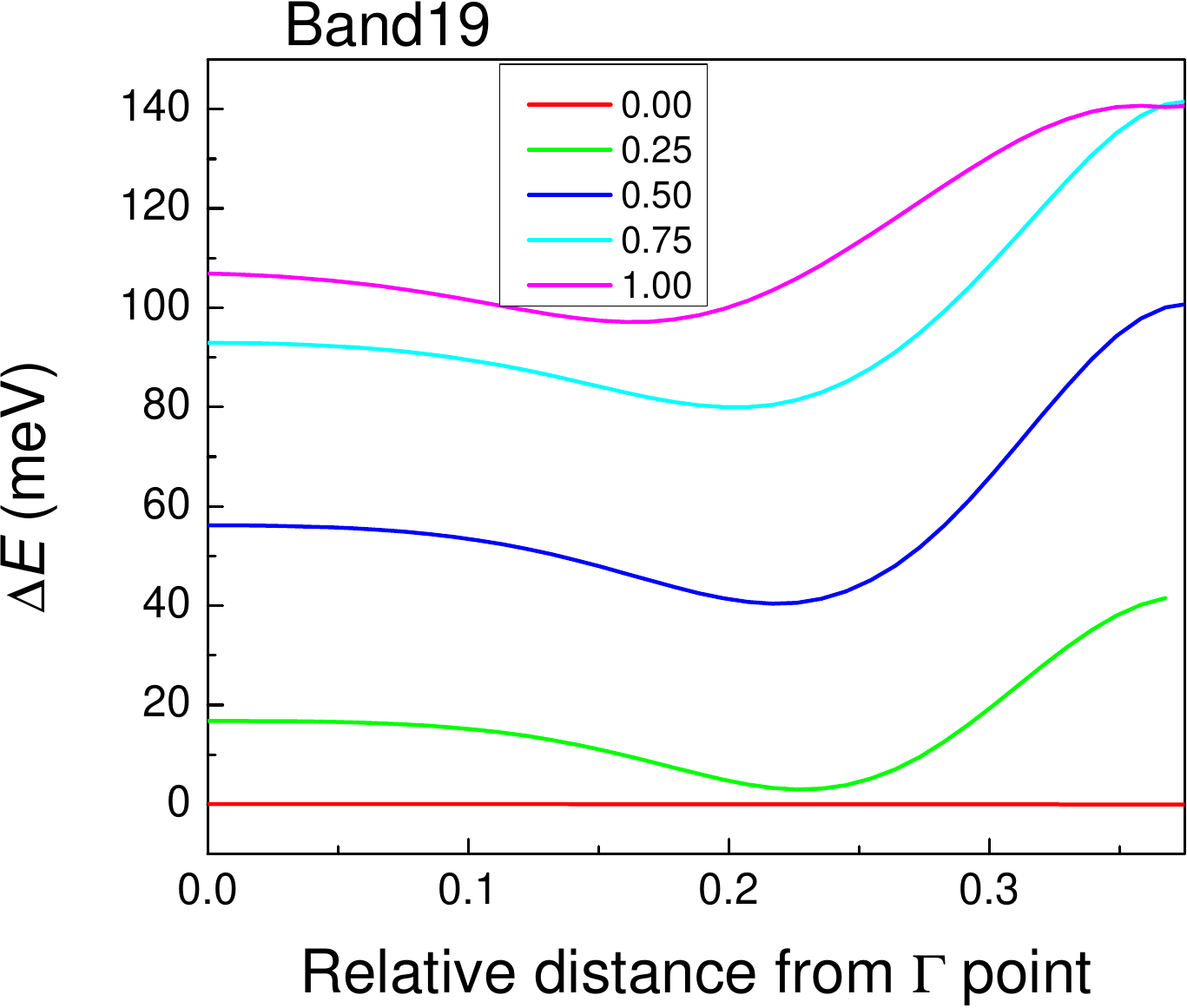}
        \caption{}
        \label{fig:de_band19}
    \end{subfigure}
    \caption{$\Delta_n$ along the $R$--$\Gamma$--$\bar {R}$ segment, measured from $\Gamma$, for the ligand-arrangement series interpolating between the $P\bar1$ ($0\%$) structure and $P1$ ($100\%$) \ce{O} positions. (a) With the $P\bar1$ ($0\%$) structure, bands $9$, $14$, $18$, and $19$. (b) The same four bands in the $P1$ structure ($100\%$). (c) Band $14$ across the full series ($0\%,25\%,50\%,75\%,100\%$). (d) Band $19$ across the same series.}
    \label{fig:de_all}
\end{figure*}

With the ligands repositioned to restore the inversion centre [Fig.~\ref{fig:p1}(b)], the picture collapses: the two \ce{Mn} moments become equal and opposite to high precision, the net moment vanishes exactly (moments for the full displacement series are tabulated in SM, Sec.~S4), and the spin-up and spin-down bands coincide throughout the zone, $\Delta_n(\mathbf{k})\simeq 0$, visible as the red dispersion curves overlying the blue ones. The exact cancellation of the net moment follows already from the pure spatial inversion relating the two \ce{Mn} (and two \ce{O}) sites: this real-space site equivalence fixes the two moment magnitudes to be equal regardless of spin direction. Cancellation of the bands at every $\mathbf{k}$ is a stricter requirement. A pure spatial inversion exchanges the two magnetic sublattices and so is not, by itself, a symmetry of the magnetic Hamiltonian; the protecting operation is the antiunitary composite $\mathcal{PT}$, inversion combined with time reversal (Sec.~\ref{sec:discussion}). Because the two sublattices are converged independently in the self-consistent calculation, this $\mathcal{PT}$ relation is not enforced numerically, and Fig.~\ref{fig:de_all}(a) shows the consequence directly: a residual splitting below $1$~meV survives in every band inspected, roughly two orders of magnitude smaller than in the original arrangement [Fig.~\ref{fig:de_all}(b)] but reproducing the same pattern of signs between bands. Because the two opposite-spin sublattices must be treated as distinct self-consistent magnetic components in the calculation, the exact \(PT\) relation is not imposed numerically even when the atomic positions recover inversion symmetry. The resulting sub-meV splitting should therefore be regarded not as a physical lifting of \(PT\) degeneracy, nor as random numerical noise, but as a systematic residual of the unconstrained self-consistent solution. Its preservation of the same band-dependent sign pattern as the P1 structure is consistent with this interpretation.

This reversal is the crux, and it runs opposite to the way ligands enter the altermagnetic account. In a rutile or perovskite altermagnet the ligand octahedra are what \emph{create} the symmetry that relates the sublattices: the metal sites alone do not realize the connecting rotation, and only the surrounding oxygen arrangement supplies it. Here the ligands do the opposite. Repositioning them does not destroy a connecting symmetry but \emph{restores} one --- an inversion relating the two now-equivalent \ce{Mn} sites --- which connects the opposite-spin sublattices at fixed $\mathbf{k}$ and re-imposes the Level-I degeneracy of Eq.~\eqref{eq:translation_degeneracy}, up to the numerical residual just discussed. In both readings the ligand arrangement is what controls the symmetry: it can build a connecting operation where the bare sublattices have none, or restore one that the original arrangement had suppressed. The only change between the two calculations is the ligand arrangement, and with it the splitting appears and collapses. The splitting of panel (a) is therefore not a consequence of the crystallographic symmetry, which is trivial (or, in panel (b), centrosymmetric but magnetically unprotected) in both cases, but of the anisotropic arrangement of the \ce{Mn}--\ce{O} network. This is the triclinic counterpart of the monoclinic negative control of Sec.~\ref{subsec:negative_control}: there the arrangement was stripped from a rotation/mirror-bearing cell and the altermagnetism was lost; here it is displaced, continuously, from a symmetry-free cell toward an $P\bar1$ structure, and the splitting itself collapses with it.

The removal experiment does more than confirm that the arrangement matters; it separates two roles of the ligands that are ordinarily merged. The ligands are known to set the symmetry, and in most altermagnets this is the role they are assigned: the oxygen octahedra supply the rotation that relates the sublattices, and the splitting is then credited to that symmetry. Read this way, the ligands act only through the symmetry they induce, and ``arrangement'' would be no more than a route to a symmetry that does the real work. The $P1$ result blocks this reduction. Repositioning the oxygens does not remove a splitting-enabling symmetry --- there is none to remove --- yet the splitting collapses. The ligands therefore act on the splitting by a route that does not pass through any sublattice-relating symmetry: they set the directional, sublattice-dependent orbital occupation and hopping that make the two spin channels differ, and it is the loss of this, not the loss of a symmetry, that closes the gap. The role of the ligand arrangement is thus not reducible to symmetry induction. It induces the symmetry, and, independently, it generates the splitting; the second role is the origin, and it is the one the symmetry-centered account leaves implicit.

Because the two ligand arrangements are the endpoints of a continuous displacement, the same calculation can be repeated at intermediate points, tracing out how individual bands respond as the arrangement moves between them. Figures~\ref{fig:de_all}(c) and~\ref{fig:de_all}(d) follow two representative bands along this series. Band $14$ [panel (c)] carries a zero crossing on this section, absent from bands $9$, $18$, and $19$: as the arrangement moves from the $P\bar1$ endpoint toward the original one, the crossing migrates continuously rather than appearing at a fixed location, and the magnitude of the splitting varies non-monotonically along the way. Band $19$ [panel (d)], by contrast, develops no crossing on this section at any point in the series; its splitting keeps a single sign throughout and instead grows monotonically as the arrangement moves away from the $P\bar1$ endpoint. The two bands thus realize, concretely and continuously, the two possibilities left open by the minimal model of Sec.~\ref{subsec:P1_model}: whether $\Delta E(\mathbf{k})=0$ has a solution on a given section depends on the particular hopping coefficients, so a node's existence, not only its location, is set by the arrangement rather than by any symmetry, which here supplies none at all. Two further bands along the same series --- one developing no node at all, the other a node that is present only over part of the series and vanishes entirely beyond some intermediate point --- are given in SM, Sec.~S6, confirming that this behavior is not confined to the two bands shown here.

\subsection{Why the surface is unavoidable: a minimal model}
\label{subsec:P1_model}

The necessity behind the calculation is seen in a minimal tight-binding model that keeps only what matters: the directional, sublattice-dependent hopping that an anisotropic arrangement produces. With spin and lattice decoupled in the nonrelativistic limit, the two spin channels disperse independently. Taking one orbital per site on a simple lattice, and writing $\varepsilon_\sigma(\mathbf{k})$ for the spin band derived from sublattice $\sigma$, the splitting is
\begin{equation}
    \Delta E(\mathbf{k})
    =
    \varepsilon_\uparrow(\mathbf{k})-\varepsilon_\downarrow(\mathbf{k})
    =
    2\sum_{i=x,y,z}(t_{1i}-t_{2i})\cos k_i,
    \label{eq:tb_split}
\end{equation}
where $t_{1i}$ and $t_{2i}$ are the hopping amplitudes along direction $i$ on the two sublattices. Only the cosine appears because $E_\sigma(-\mathbf{k})=E_\sigma(\mathbf{k})$ in the nonrelativistic limit; the same evenness that forbids an odd-parity $p$-wave splitting fixes the form of Eq.~\eqref{eq:tb_split}.

Two features follow with no further assumption. First, when the arrangement is anisotropic the hoppings differ between sublattices, $t_{1i}\neq t_{2i}$, and $\Delta E(\mathbf{k})=0$ is one equation in the three components of $\mathbf{k}$; whether it has a solution depends on the full set of coefficients, and a sufficiently large uniform offset between the sublattices could in principle leave $\Delta E(\mathbf{k})$ of one sign throughout the zone.  A nodal surface is thus accidental in existence as well as in location: nothing in the point group $1$ forces one to appear.  What symmetry can force is only a conditional statement: a monoclinic crystal supplies a single enforced plane (Sec.~\ref{subsec:tiers}), and since a $p$-wave splitting is forbidden by $\Delta_n(-\mathbf{k})=\Delta_n(\mathbf{k})$, a monoclinic state that is altermagnetic at all must carry a second, unenforced surface for its four-sector texture to exist --- accidental, yet unavoidable given that the texture is realized.  No such texture constrains the triclinic case, so there a zero of $\Delta E(\mathbf{k})$, where one occurs, is doubly accidental: unforced in existence and, second, unforced in location, set by the coefficients $t_{1i}-t_{2i}$ --- by the arrangement --- and by nothing else.  No operation pins it to a fixed plane or relates it to a partner.

Imposing a symmetry does exactly, and only, the latter. A spin-exchange operation that interchanges the $x$ and $y$ directions constrains the hoppings, $t_{1x}=t_{2y}$ and $t_{1y}=t_{2x}$, and Eq.~\eqref{eq:tb_split} collapses to
\begin{equation*}
    \Delta E(\mathbf{k})
    =
    2(t_{1x}-t_{1y})(\cos k_x-\cos k_y),
    \label{eq:tb_collapse}
\end{equation*}
whose zeros are the paired planes $k_x=\pm k_y$. The two factors carry the two roles cleanly. The form $(\cos k_x-\cos k_y)$ is supplied by the symmetry: it is what pairs one plane with its partner, so that once the arrangement fixes one nodal plane the symmetry replicates it at the symmetry-related orientation. The coefficient $(t_{1x}-t_{1y})$ is supplied by the arrangement: it sets whether the splitting exists and how large it is. Were the arrangement isotropic in the plane, $t_{1x}=t_{1y}$, the splitting would vanish identically even with the symmetry fully present. Symmetry does not create the nodal structure and does not, on its own, guarantee any splitting; it takes the surface the arrangement has already produced and pins and replicates it into planes of definite orientation.

\subsection{The base tier of the hierarchy}
\label{subsec:P1_tier}

These results place the triclinic case at the base of the hierarchy rather than outside it. The number of enforced nodal planes decreases across the crystal classes from two (orthorhombic) to one (monoclinic) to zero (triclinic), while the accidental surfaces, generated by the orbital arrangement, persist to the bottom. The triclinic tier is the limit in which organization by symmetry has vanished entirely and only the arrangement-generated splitting remains. Including it makes the classification internally consistent: the same arrangement-generated splitting that supplies the second, accidental surface of a monoclinic altermagnet is the \emph{whole} of the splitting in a triclinic ferrimagnet, where no enforced surface accompanies it. The same accidental surfaces are not confined to the low-symmetry tiers: being generated by the arrangement rather than by symmetry, they may accompany the enforced planes in crystals of higher symmetry as well, where they are merely not required to complete the texture.

This clarifies what the accidental surface is. It is not a by-product of the enforced plane, nor a feature that the altermagnetic texture generates to complete itself. It is present in $P1$, where there is no enforced plane and no altermagnetism for it to complete, produced by the orbital arrangement alone. The monoclinic altermagnet inherits this same arrangement-generated surface and adds to it a single symmetry-enforced plane; the orthorhombic altermagnet adds two. Read in this direction, from the base tier upward, the causal order is the physical one. The orbital arrangement generates spin splitting and its accidental nodal surfaces whenever it is anisotropic, with or without symmetry. Symmetry, when present, does not generate the splitting but organizes it --- fixing the location and enforcing the existence of particular nodal planes, and thereby converting an otherwise unorganized splitting into an altermagnetic texture of definite wave character. The triclinic $P1$ ferrimagnet is the case in which this organizing role is absent, and it demonstrates, more directly than any altermagnet can, that the splitting itself does not require it.

\section{Discussion}
\label{sec:discussion}
The hierarchy of the preceding sections was built to answer a question of symmetry: given a magnetic decoration and a crystal, is spin splitting admissible, which operations relate the spin sectors, and where do they enforce nodes. Carried to the triclinic base, the same construction answers a different and larger question --- what produces the splitting in the first place --- and the answer is not symmetry. We collect that conclusion here and state its consequences.

Throughout this work we consider the spin--orbit-free collinear limit, so that spin is a separate label and the object of interest is the same-$\mathbf{k}$ spin degeneracy $E_{n\uparrow}(\mathbf{k})=E_{n\downarrow}(\mathbf{k})$ and its lifting.  When it holds, this degeneracy is protected by an antiunitary composite such as $\mathcal{PT}$ or $\mathcal{T}\mathbf{t}$ --- time reversal combined with inversion or with a sublattice-exchanging translation.  Its lifting is what we call the spin splitting throughout.

Seen this way, the lifting of spin degeneracy in a compensated collinear magnet is not the exclusive signature of altermagnetism. It occurs whenever no antiunitary composite protects the degeneracy. An altermagnet is the case in which $\mathcal{PT}$ and $\mathcal{T}\mathbf{t}$ are absent but time reversal combined with a rotation or mirror survives, protecting the degeneracy on particular momentum manifolds and leaving it lifted elsewhere --- the enforced nodal planes. The triclinic $P1$ ferrimagnet is the case in which no such composite exists at all: with only the identity, there is no operation for time reversal to combine with, so nothing protects the degeneracy at any momentum. That spin splitting appears is then unremarkable from a symmetry standpoint --- expected, even, since nothing forbids it. This very ordinariness is the point. The triclinic ferrimagnet has been passed over precisely because its splitting seemed too obvious to be worth examining, and in passing it over the field assumed that a state which cannot be an altermagnet has no spin splitting to account for. It does, and accounting for it exposes what has been doing the work all along.

What has been doing the work is the anisotropic arrangement of the magnetic orbitals. Its role is visible once the crystal classes are read as a single axis rather than as a list with the triclinic entry deleted. Across that axis the number of symmetry-enforced nodal planes falls from two in the orthorhombic case to one in the monoclinic to none in the triclinic, while the arrangement-generated splitting, and the accidental surfaces that accompany it, persist unchanged to the base. The compensated spin splitting is even under $\mathbf{k}\to-\mathbf{k}$ in every case, an odd-parity $p$-wave form being forbidden, so it always requires a sufficient set of nodal surfaces; symmetry supplies part of that set as enforced planes and the arrangement supplies the rest as accidental surfaces. As the enforced count falls along the axis, the accidental surfaces take up the whole burden, until at the triclinic base they carry it alone. Altermagnet and fully compensated ferrimagnet are the two ends of this one axis, not two unrelated phenomena: the former is the arrangement-generated splitting with enforced planes added and organized into a definite wave character, the latter the same splitting with no enforced plane and no organization, its accidental surfaces multiple and band-dependent.

The arrangement is therefore prior to symmetry in two distinct senses, and the triclinic and admissibility results establish each. It generates the splitting: with the arrangement anisotropic the splitting is present, and altering it collapses it, whether or not any symmetry is nominally retained.  This is shown most directly in Sec.~\ref{sec:triclinic}, where repositioning the same ligands --- unchanged in number and species --- to restore an inversion centre isolates the arrangement as the sole variable; the cruder test of removing the ligands outright in the monoclinic minimal model (Sec.~\ref{sec:minimal_model}) points to the same conclusion, though there several properties of the cell change at once. The minimal model of Eq.~\eqref{eq:tb_split} shows why this is generic rather than particular: an anisotropic difference in the hopping network gives one equation in three momentum components, whose regular solution set, if it exists, is generically a nodal surface, with or without any symmetry. The same model also makes precise why, when a veto operation and a rotation coexist, as in Sec.~\ref{sec:admissibility}, it is always the veto that wins.  A translation or inversion relating the two sublattices forces their local hopping environments to coincide, $t_{1i}=t_{2i}$ for every direction $i$, so the coefficient in Eq.~\eqref{eq:tb_split} vanishes identically and $\Delta E(\mathbf{k})\equiv0$ at every $\mathbf{k}$; no rotation-organized angular factor $\cos k_i$ can restore a splitting from a coefficient that is exactly zero.  This is the minimal-model version of the \ce{KV2Se2O} and \ce{KCuF3} result: the rotation relating opposite-spin sublattices is present in every case, yet only where the veto is absent does it have a nonzero splitting to organize. And it determines the symmetry itself: the operations that relate the opposite-spin sublattices are not fixed by the magnetic ions alone but by the arrangement of the surrounding ligand octahedra, which can build a connecting rotation where the bare sublattices have none, or, as in $P1$-MnO, whose removal revives a connecting inversion that the arrangement had suppressed. The same structural anisotropy also fixes admissibility, through the octahedral tilting that removes or retains the parent-restoring translation of the Level-I veto (Sec.~\ref{sec:admissibility}). At each level of the hierarchy --- whether splitting is admitted, which symmetry organizes it, how large it is --- the arrangement is the determining input and symmetry is its consequence. This dissolves the apparent question of whether arrangement or symmetry comes first: symmetry is not an independent given but is set by the arrangement, which then also generates the splitting that the symmetry goes on to organize. That the two roles are genuinely separate, and not one role seen twice, is established most cleanly by the ligand-repositioning experiment of Sec.~\ref{sec:triclinic}: repositioning the oxygens, without changing their number or species, does not lower the symmetry --- it raises it --- yet the splitting collapses, so the arrangement must act on the splitting by a route that does not pass through the sublattice-relating symmetry.  The same conclusion is corroborated, less surgically, by the ligand-removal test of the monoclinic minimal model (Sec.~\ref{sec:minimal_model}).

This reading also places the present results in relation to recent work that has loosened the tie between spin splitting and crystal symmetry from other directions. Interaction-driven orbital ordering has been shown to produce altermagnetic splitting in a lattice without crystallographic sublattice anisotropy \cite{Leeb2024}, and a locally assembled common point-group symmetry can support it in amorphous solids that lack global rotational symmetry \cite{Ornellas2026}. In both, an orbital ordering or arrangement must be established before the organizing symmetry can act; the arrangement is, in effect, already prior in those constructions. What is left implicit there, and what the triclinic limit makes unavoidable, is that this priority is causal rather than incidental: the arrangement is the origin of the splitting, and the symmetry it revives only organizes a splitting that is already present. Made explicit for a case with no organizing symmetry at all, the statement cannot be reabsorbed into a symmetry criterion --- and it bears directly on those realizations, since in each, had the arrangement revived no relating symmetry, the splitting would not have vanished but merely gone unorganized.

\subsection{Why symmetry organizes but does not generate the splitting}
\label{subsec:Sym-Org}
The preceding paragraphs have pressed the accumulated results as far as they will go.  The triclinic ferrimagnet shows a splitting present with no symmetry at all, and removed when the arrangement is; the ligand-removal experiment separates the generative role of the arrangement from the symmetry-inducing one; and the minimal model shows, down to a single coefficient, why a coexisting veto always overrides a coexisting rotation.  This is as strong a case for the arrangement as the assembled results can make, and as an argument it is still incomplete.  None of it independently rules out that a rotation or mirror, where one is present, might generate a splitting of its own alongside the one the arrangement produces.  Without settling that point, the claim that the arrangement comes first is a reading of the evidence, not a conclusion forced by it.  What is required is the stronger statement that symmetry cannot generate the splitting at all, and this can be shown directly.

The standard description of altermagnetic splitting already contains the reason, once its logic is examined. One says that a band which is spin-up at $\mathbf{k}$ becomes spin-down at $R\mathbf{k}$, and spin-down at $\mathbf{k}$ becomes spin-up at $R\mathbf{k}$, and reads this as the splitting. But to interchange the spin-up and spin-down bands at $\mathbf{k}$ presupposes that the two are already distinct there --- already split. If they were degenerate at $\mathbf{k}$, as in a conventional antiferromagnet, the rotation would interchange two equal energies and leave them degenerate at $R\mathbf{k}$; nothing would be produced. A rotation relating opposite-spin sublattices can only interchange bands that are already separated in energy. It cannot separate them. The admissibility results of Sec.~\ref{sec:admissibility} are the concrete demonstration: in A-type \ce{KV2Se2O} and \ce{KCuF3} a spin-exchanging screw does relate the sublattices, so the rotational condition for altermagnetism is met, yet no splitting appears, because a parent-restoring translation holds the two spin channels degenerate and leaves the rotation nothing to interchange. Where a symmetry is available to relate the sublattices, it can reverse a splitting between $\mathbf{k}$ and $R\mathbf{k}$, but it cannot bring one into being.

Two further facts already established in this hierarchy illustrate the same logical point rather than standing as separate curiosities. The first is why admissibility must be resolved before organization: Sec.~\ref{sec:admissibility} showed that in A-type \ce{KV2Se2O} and \ce{KCuF3} the rotation required to organize a splitting is present in full, yet a coexisting translation veto holds the two spin channels degenerate regardless of it.  If the rotation could generate a splitting on its own, its presence would guarantee one irrespective of the veto; that it does not confirms the rotation's role is confined to organizing a splitting that, when the veto holds, is simply not there to act on.  The second is the zone-boundary self-invariance of C-type \ce{LaMnO3} (Sec.~\ref{subsec:AGC}): there, a path is forced nodal not by the operation that relates its own endpoints but by an unrelated partner operation whose invariant plane the path happens to lie on --- symmetry imposing a zero without regard to what that path's own relation would suggest.  In neither case does symmetry weigh whether a splitting should exist and act accordingly.  It either forces a zero, as the veto and the zone-boundary plane both do, or reverses a sign, mechanically, on whatever splitting is or is not already present.  The formal argument below states the same fact in general terms.

The same conclusion follows formally, without reference to any hopping or ligand mechanism, from the symmetry relation alone.  Suppose the net moment vanishes and the two spin sublattices are related by a fourfold rotation $C_4$ and by no translation or inversion.  The only constraint this places on the bands is
\begin{equation*}
E_\uparrow(\mathbf{k}) = E_\downarrow(C_4\mathbf{k}).
\end{equation*}
This relation does not imply a splitting.  It is satisfied equally in two ways.  If $E_\uparrow$ is not invariant under $C_4$, then $E_\downarrow(\mathbf{k})=E_\uparrow(C_4^{-1}\mathbf{k})\neq E_\uparrow(\mathbf{k})$ and the bands are split; but if $E_\uparrow$ is invariant under $C_4$, the same relation gives $E_\downarrow(\mathbf{k})=E_\uparrow(\mathbf{k})$ and the bands are degenerate at every momentum.  The rotational relation is thus compatible with a splitting and with its complete absence, and what selects between them is whether $E_\uparrow(\mathbf{k})$ is anisotropic under $C_4$ --- a property of the orbital arrangement on each sublattice, not of the rotation that relates them.  The two symmetry conditions, a vanishing net moment and a rotation exchanging the sublattices, therefore permit a splitting without generating one; the anisotropy that turns the permission into an actual splitting is supplied by the arrangement.

This fixes what symmetry does, and it is not one thing but two, neither of them generation. On its invariant manifolds --- the rotation axes, mirror and glide planes --- a spin-exchanging operation forces $\Delta_n=0$, pinning the enforced nodal planes; and away from them it reverses the sign of an already-present splitting between $\mathbf{k}$ and $R\mathbf{k}$, arranging it into the alternating $d$-, $g$-, or $i$-wave pattern. Both roles act on a splitting that must already exist. The splitting itself has a single origin: a difference between $E_\uparrow(\mathbf{k})$ and $E_\downarrow(\mathbf{k})$ at a generic momentum, which requires the two sublattices to be electronically inequivalent, which is a property of the arrangement. This completes the reordering the paper has argued for. The anisotropic arrangement of the magnetic orbitals is the origin of the nonrelativistic spin splitting and of its nodal surfaces; symmetry, when present, organizes that splitting --- pinning its enforced nodes and reversing its sign across momentum space --- and, through the admissibility veto, decides whether it is organized at all, but it does not produce it. The altermagnet is what the arrangement-generated splitting becomes when a symmetry is present to organize it, and the fully compensated ferrimagnet is what it remains when none is.

Throughout, this separation has relied on real, time-reversal-symmetric hopping within each spin channel, whose even parity, $E_\sigma(-\mathbf{k})=E_\sigma(\mathbf{k})$, is what excludes an odd-parity ($p$-wave) splitting alongside the even $d$-, $g$-, $i$-wave family treated here. Recent work has shown that breaking this residual time-reversal symmetry as well, for instance through loop currents encoded as complex hoppings, lifts that exclusion and permits an odd-parity texture in an otherwise collinear magnet \cite{Lin2025,Leeb2026}. The role of the loop current there is exactly the veto's role here, only in reverse: where a translation or inversion vetoes any splitting by enforcing $E_\sigma(-\mathbf{k})=E_\sigma(\mathbf{k})$, the loop current clears an analogous veto against the odd-parity form by breaking that same relation. Either way, imposing or clearing the condition only decides what is symmetry-admissible; it supplies no amplitude. We expect the same separation to hold there: the loop current only permits a $p$-wave form, while the amplitude of any resulting splitting would still require a real, sublattice-anisotropic hopping network of the kind identified in this work. Establishing this explicitly is a natural extension we leave for future work.

\section{Summary and outlook}
\label{sec:conclusion}

We have set out a systematic way to decide how altermagnetism arises in a given collinear magnet, resolving in order the admissibility of spin splitting, the network of sublattice-relating operations, and the enforced nodal geometry.  Applying this hierarchy across orthorhombic, monoclinic, and triclinic crystals, we found that the splitting does not originate in the crystal symmetry but in the anisotropic arrangement of the magnetic orbitals and their ligands; symmetry organizes an already-present splitting, and does not produce it.  This rests on two independent grounds.  Empirically, triclinic $P1$-\ce{MnO} is spin split with no symmetry relating its sublattices, and the splitting collapses by two orders of magnitude when the same ligands are repositioned to restore an inversion centre. That this had gone unexamined is itself telling.  Three ordinary expectations discouraged it: that a ferrimagnet is uncompensated and its splitting unremarkable, that a compensated magnet is either an antiferromagnet or, with the right symmetry, an altermagnet, and that a triclinic splitting is a trivial consequence of having no symmetry to prevent it.  Each, taken alone, made the fully compensated ferrimagnet seem not worth a closer look, and together they left it aside --- precisely the case in which the arrangement acts alone and its role becomes visible.

We close with a suggestion the results invite but do not establish. Throughout we have kept the fully compensated ferrimagnet terminologically distinct from the altermagnet, as the standard definition requires: the one has no symmetry relating its sublattices, the other a rotation or mirror.  Yet across the crystal classes the enforced nodal planes fall from two to one to zero while the arrangement-generated splitting persists to the base, so the two might be more naturally regarded as one family --- the fully compensated ferrimagnet included not as a separate phase but as the limiting tier of altermagnetism, in which the organizing symmetry has vanished and only the arrangement-generated splitting remains, much as a square is a special rectangle rather than a separate shape.  Such a regrouping is not ours to declare; it is a matter of convention and of consensus in the community.  We note only that it would make the classification internally consistent, and that it carries a practical consequence: the symmetry criterion for altermagnetism would then have to be applied with care --- conditioned on whether the crystal retains any sublattice-relating operation at all --- rather than as a single blanket test.

\begin{acknowledgments}
This work was supported by the Basic Science Research Program through the National Research Foundation (NRF) of Korea funded by the Ministry of Education (Grant No. 2020R1I1A1A01070231).
\end{acknowledgments}

\end{document}